\documentclass[12pt,a4paper]{article}

\usepackage{geometry}
\usepackage[utf8]{inputenc}
\usepackage[linesnumbered, ruled, vlined]{algorithm2e} % Algorithm
\usepackage{amsthm}   % added,must be in first ones o.w. error: openbox is defined.
\usepackage{amsmath}
\usepackage{graphicx,psfrag,epsf}
\usepackage{enumerate}
\usepackage{mathtools}
\usepackage{amsfonts} % for \mathbb
\usepackage{bbm}      % for indicator function
\usepackage{xcolor}   % use color in text
\usepackage{verbatim}  % comment
\usepackage{natbib}  % to use \citep 
\usepackage[colorlinks=true, linkcolor=cyan]{hyperref} % link color
\usepackage[nameinlink,noabbrev]{cleveref}
\usepackage{amsmath}
\usepackage{subfiles} % enable compiling subfiles
\usepackage{comment}
\usepackage{subcaption}

\usepackage{times}
\usepackage{bm}
\usepackage{natbib}

\usepackage{amsfonts} % to use \mathbb
\usepackage{NewCommand}      % contains self-defined newcommands
\usepackage{bbm}
\usepackage{comment}
\usepackage{caption}

\usepackage[perpage]{footmisc} %to use more footnotes

\title{On the error control of invariant causal prediction}
\author{Jinzhou Li, Jelle J Goeman}
\date{\today}

\begin{document}
\maketitle

\begin{abstract}

%Invariant causal prediction (ICP, \cite{peters2016causal}) provides a useful framework for identifying causal predictors of a response using heterogeneous data from multiple environments. One valuable property of ICP is that it guarantees no false causal discoveries with high probability. Such a guarantee, however, can be overly conservative in some applications, resulting in few or no causal discoveries. This raises a natural question: can we equip ICP with less conservative error guarantees and thereby extract more causal information from the data? In this paper, we address this question by focusing on two widely used and more liberal guarantees: false discovery rate control and simultaneous true discovery bounds. A key step in our approach is to reformulate the original ICP procedure as a multiple testing problem. We then adopt the recently proposed e-Closure principle to obtain (simultaneous) false discovery rate control for ICP, together with new p-to-e calibrators tailored to this setting. We also derive simultaneous true discovery bounds via closed testing, which provide additional causal information for free, in the sense that no extra assumptions are required and all discoveries from the original ICP procedure are retained. Through simulations and a real data application on educational attainment of teenagers in the US, we show that these more liberal guarantees can improve the practical usefulness of ICP.

%% JRSSB require no citations and abbreviations in abstract
Invariant causal prediction provides a useful framework for identifying causal predictors of a response using heterogeneous data from multiple environments. One valuable property of the original invariant causal prediction method is that it guarantees no false causal discoveries with high probability. Such a guarantee, however, can be overly conservative in some applications, resulting in few or no causal discoveries. 
This raises a natural question: can invariant causal prediction be equipped with less conservative error guarantees and thereby extract more causal information from the data? In this paper, we address this question by focusing on two widely used and more liberal guarantees: false discovery rate control and simultaneous true discovery bounds. A key step in our approach is to reformulate invariant causal prediction as a multiple testing problem. We then adopt the e-Closure principle to obtain (simultaneous) false discovery rate control, together with new p-to-e calibrators tailored to this setting. We also derive simultaneous true discovery bounds via closed testing, which provide additional causal information without requiring extra assumptions and retain all discoveries from the original invariant causal prediction method. Through simulations and a real data application on educational attainment of teenagers in the United States, we show that these more liberal error control guarantees can improve the practical usefulness of invariant causal prediction.

\end{abstract}


% \tableofcontents

%%%%%%%%%%%%%%%%%%%%%%%%%%%%%%%%%%%%%%%%
\newpage
\section{Introduction} \label{sec:intro}

Discovering causal predictors of a response of interest is one of the principal goals of scientific research. However, based solely on observational data, causal relationships may be unidentifiable or difficult to estimate. In such cases, heterogeneous data from different environments can enhance identifiability and provide additional information, and are therefore a valuable resource for causal discovery.

In the multi-environment setting, \cite{peters2016causal} proposed a useful method called invariant causal prediction (ICP), which identifies causal predictors by exploiting the invariance of the conditional distribution of the response given its causal predictors across environments \citep{wright1921correlation, haavelmo1944probability, aldrich1989autonomy, hoover1990logic, dawid2010identifying, scholkopf2012causal}. 
Specifically, ICP generates p-values $p_S$ for each set $S \subseteq [m]=\{1,\dots,m\}$, where $m$ is the number of variables, by testing whether $S$ is invariant across environments (see \eqref{def:H_0_S}). The estimated set of causal predictors, $\whS^{\text{ICP}}$ (see \eqref{Shat:ICP}), is then obtained based on these p-values.

%Compared to other causal discovery methods, 
One notable advantage of ICP is that it guarantees, with high probability, that all its discoveries are true causal predictors:
\begin{align*}
P \left( \whS^{\text{ICP}} \subseteq S^* \right)=
P \left( |\whS^{\text{ICP}} \setminus S^*|=0 \right)
\geq 1-\alpha,
\end{align*}
where $\alpha \in (0,1)$ is a nominal level and $S^*$ denotes the set of true causal predictors.
%, and $\mN^*= [m] \setminus S^*$ denotes the set of null causal predictors.
This guarantee is known as familywise error rate (FWER) control.

Controlling the FWER, however, can be overly conservative in some applications, particularly when causal predictors are highly correlated with non-causal ones, as noted in the literature \citep{heinze2018invariant, fan2024environment}.
In such cases, ICP may result in few or no causal discoveries. 
%resulting in minimal useful causal information.
This raises a natural question: is it possible to control more liberal error rates for ICP, thereby extracting more causal information?
Two of the most widely used and less conservative alternatives to FWER control are false discovery rate (FDR) control \citep{benjamini1995controlling, benjamini2001control} and simultaneous true discovery bounds \citep{genovese2006exceedance, goeman2011multiple}. We therefore consider the following two problems based on the p-values $p_S$ generated by ICP:
\begin{itemize}
\item[1.] \textbf{FDR control:} Can we obtain a discovery set $\whS$ satisfying
\begin{align*}
\bbE \left[ \frac{|\whS \setminus S^*|}{\max\{1, |\whS|\}} \right] \leq \alpha?
\end{align*}
In particular, since FWER control implies FDR control (so $\whS^{\text{ICP}}$ already guarantees FDR control), the goal here is not merely to obtain a set with FDR control, but to obtain more discoveries than $\whS^{\text{ICP}}$ whenever possible.
\item[2.] \textbf{Simultaneous true discovery bound:} Can we obtain a function $t_{\alpha}: 2^{[m]} \rightarrow \{0,1,\dots,m\}$ such that
\begin{align*}
P( |R \cap S^*| \geq t_{\alpha}(R) \text{ for all } R \subseteq [m]) \geq 1-\alpha?
\end{align*}
The bound $t_{\alpha}(R)$ enables users to examine any set $R$ they consider interesting, which may be constructed using machine learning methods or domain knowledge, and informs users, with high probability, that at least $t_{\alpha}(R)$ true discoveries are contained in $R$.
\end{itemize}

We note that the simultaneous true discovery bound $t_{\alpha}(R)$ is equivalent to a corresponding simultaneous false discovery bound, in the sense that from $t_{\alpha}(R)$ we obtain the simultaneous false discovery bound $|R| - t_{\alpha}(R)$, which satisfies
\begin{align*}
	P( |R \setminus S^*| \leq |R| - t_{\alpha}(R) \text{ for any } R \subseteq [m]) \geq 1-\alpha.
\end{align*}
We focus on the simultaneous true discovery bound throughout this paper because it offers a more convenient interpretation for applications. Many other error rates, such as k-FWER \citep{LehmannRomano2006KFWER, sarkar2007stepup} and false discovery exceedance \citep{dudoit2004multiple, korn2004controlling, romano2006stepup}, can be viewed as special cases of the simultaneous true discovery bound; see \cite{goeman2021only} for further discussion.
% In particular, it  is desirable that setting $R=\whS^{\text{ICP}}$ yields $t_{\alpha}(R) = |R|$, which recovers the FWER control of $\whS^{\text{ICP}}$.

\citet{heinze2018invariant} also investigated how to mitigate the conservativeness of FWER control in ICP and proposed the notion of defining sets. A defining set is the smallest set guaranteed to contain at least one true causal predictor, making it particularly useful when ICP returns an empty set. Such defining sets arise naturally from the closed testing perspective that we adopt \citep{goeman2011multiple}.

\subsection{Our contributions}
We address the above two problems by proposing new methods for controlling the FDR and obtaining simultaneous true discovery bounds in ICP.

% To address these two goals, 
A key step of our approach is to formulate a multiple testing problem (see \eqref{def:direct-hypothesis-i}) that directly tests whether causal predictors are null, and to construct p-values for these hypotheses based on the $p_S$ values generated by ICP through invariance testing.
Based on this formulation, we propose valid p-values $p_i^*$ for a single hypothesis (see~\eqref{p-value-direct-approach-i}) and $p_S^*$ for an intersection hypothesis (see~\eqref{p-value-direct-approach-S}).

It may be surprising that, without any multiple testing correction, the discovery set obtained by directly comparing $p_i^*$ to the significance level recovers the original ICP method, and thus already controls both the FWER and the FDR. 
Moreover, we show that $p_i^*$ is optimal when the underlying invariance tests in ICP are optimal, indicating that this construction is not conservative. 
Together, these observations suggest that obtaining non-trivial FDR control for ICP via p-values is challenging.
To address this, we adopt the recently proposed e-Closure principle \citep{xu2025bringing} to bring FDR control to ICP. 
Specifically, we construct an e-collection from ICP by transforming the p-values into e-values via a carefully designed p-to-e calibrator, and then apply the e-Closure principle to obtain FDR control. 
Notably, this approach yields simultaneous FDR control, meaning that the guarantee holds for multiple discovery sets simultaneously, thereby allowing users to choose among FDR-controlled sets for further investigation. 
Empirical results on both simulated and real data confirm that this approach yields non-trivial FDR control, in the sense that it can lead to more discoveries than the original ICP method with FWER control.

For the second problem introduced above, namely simultaneous true discovery bounds, the multiple testing formulation allows us to readily apply the closed testing procedure \citep{marcus1976closed, goeman2011multiple}. Importantly, we show that, due to the special construction of $p_i^*$, the closed testing adjustment step is unnecessary, which substantially reduces the computation by avoiding the need to carry out all the steps of the standard closed testing procedure. 
In particular, the resulting simultaneous bounds recover the original ICP result while generally providing additional causal insight, and thus offer a direct substitute for the original ICP method.
We also present a simultaneous bound based on the original ICP formulation and show that it is equivalent to the one obtained via closed testing. This connection is useful as it provides a clearer understanding of statistical properties such as admissibility, which are less evident in the original ICP formulation but have been well studied in the closed testing literature \citep{goeman2021only}.

These two error control guarantees for ICP go beyond simply applying a more liberal threshold to the same set of p-values used in FWER control. In particular, both approaches require testing all intersection hypotheses and can therefore extract richer causal information. 

To give a brief preview of the practical use of the proposed error criteria and their corresponding approaches, we consider a simulated dataset from Section~\ref{sec:SimuAndRealData}. To illustrate their potential, we use a relatively extreme example with nominal level $0.2$. The true causal predictors are $\{1,3,4,5,7,8,9,10\}$. 
In this case, the p-values for testing whether each variable is a null causal predictor (see \eqref{p-value-direct-approach-i}) are $0.263, 0.246, 0.263, 0.049, 0.263, 0.246, 0.263$, $0.234, 0.223,$ and $0.263$.
Most of these p-values are around $0.2$, and only variable $4$ appears significant based on the individual p-values.
The original ICP method returns a single discovery $\{4\}$, while our simultaneous FDR control procedure yields a collection of discovery sets with simultaneous FDR guarantees, including larger sets such as $\{2,4,5,6,8,9,10\}$. 
At the same time, the simultaneous bounds recover the original ICP result, as the true discovery lower bound for $\{4\}$ is one, while also providing additional insight. For example, the set $\{1,4,6,9\}$ has a true discovery lower bound of three, implying with $80\%$ confidence that at least three variables in this set are true causal predictors.

The rest of the paper is organized as follows. In Section~\ref{sec:recap}, we review the original ICP method. Section~\ref{sec:MultiFrameICP} introduces the multiple testing formulation together with the newly proposed p-values. Section~\ref{sec:subsecICPFDR} develops simultaneous FDR control for ICP via e-Closure and introduces new p-to-e calibrators. Section~\ref{sec:SimulTDICP} derives simultaneous true discovery bounds for ICP. Section~\ref{sec:SimuAndRealData} demonstrates the proposed methods through simulations and a real data application, and Section~\ref{sec:discuss} concludes with discussion and future directions.

%To give a brief preview of the practical use of the proposed error criteria and their corresponding approaches, we consider a simulated dataset from Section~\ref{sec:SimuAndRealData}, where all methods are applied at the nominal level $0.1$ and the true causal predictors are $\{1,3,4,5,7,8,9,10\}$. Here, the original ICP method returns the set $\{3,4,5\}$ as causal predictors, while our simultaneous FDR control procedure yields a collection of discovery sets with simultaneous FDR guarantees, including larger sets such as $\{1,3,4,5,10\}$. % add which are true variables 
%At the same time, the simultaneous bounds recover the original ICP result, as the true discovery lower bound for $\{3,4,5\}$ is three, while also providing additional insight. For example, the full set $\{1,2,3,4,5,6,7,8,9,10\}$ has a true discovery lower bound of six, implying with $90\%$ confidence that at least six variables are true causal predictors.

% Further simulation results and a real data application are presented in Section~\ref{sec:SimuAndRealData}.

% To summarize, our contributions are as follow:

\section{A brief recap of invariant causal prediction} \label{sec:recap}
For the sake of simplicity, we use linear models \citep{peters2016causal} for illustration.
Our methods can also be applied to non-linear cases \citep{heinze2018invariant} directly.
Consider the following linear structural equation model:
\begin{equation*}
	Y^e \leftarrow \beta_1 X^e_1 + \cdots + \beta_m X^e_m + \epsilon^e =\beta_{S^*}^T X^e_{S^*} + \epsilon^e,
\end{equation*}
where $S^* = \{ i \in [m]: \beta_i \neq 0 \} $ is the set of causal predictors of $Y$, 
$e \in \mE$ denotes one environment,
$\epsilon^e \sim F_{\epsilon}$, $\epsilon^e \independent X^e_{S^*}$, and $X^e_i$ can have arbitrary distribution for different $e$.
The goal is to estimate the set of causal predictors $S^*$.

The key idea of ICP is to test whether the conditional distribution of $Y^e$ is invariant across environments given predictors $X^e_S$ for each $S \subseteq [m]$. This leads to the following null hypothesis:
\begin{equation}\label{def:H_0_S}
	\begin{aligned} 
		H_{ 0, S }(\mE): \
		& \text{there exists a distribution $F_{\epsilon}$ and a vector } \gamma \in \bbR^m
		\text{ with } \operatorname{supp}(\gamma)\subseteq S
		% Support means that the elements in S are non-zero
		% \gamma_k = 0 \text{ if } k \notin S 
		\text{ such that } \\ 
		&\text{for all } e \in \mE,
		Y^e = \gamma^T X^e  + \epsilon^e, 
		\text{where } \epsilon^e \sim F_{\epsilon} \text{ and } \epsilon^e \independent X^e_S.
	\end{aligned}
\end{equation}
For details on testing this hypothesis, please refer to \cite{peters2016causal} and \cite{ heinze2018invariant}.

Under the assumptions that there are no latent variables and $\mE$ does not contain the environment where $Y$ is intervened,
it is clear that $H_{0, S^*}(\mE)$ is true.
However, there may be sets $S$ other than $S^*$ for which $H_{0, S}(\mE)$ is true, causing an identifiability issue.
To this end, \cite{peters2016causal} defined the so-called identifiable causal predictors under $\mE$ as follows:
\begin{align} \label{def:IdentifiableCausalPredictor}
	S(\mE) = \bigcap_{S: H_{ 0, S }(\mE) \text{ is true }} S.
\end{align}
Note that $S(\mE) \subseteq S^*$ as $H_{ 0, S^* }(\mE)$ is true.

ICP estimates the causal predictors by using the sample version of $S(\mE)$ in two steps: 
\begin{itemize}
    \item[(i)] For each $S \subseteq [m]$, test $H_{ 0, S }(\mE)$ at level $\alpha \in (0,1)$ to obtain p-value $p_S$.
    \item[(ii)] Obtain the estimated causal predictors by
\begin{equation} \label{Shat:ICP}
	\whS^{\text{ICP}}(\mE) = 
 %\underset{S: p_{S } > \alpha} {\bigcap} S.
 \underset{S: H_{ 0, S }(\mE) \text{ is not rejected }} {\bigcap} S.
\end{equation}
Unlike \cite{peters2016causal}, where $\whS^{\text{ICP}}(\mE)$ is defined as the empty set when all $H_{0,S}(\mE)$ are rejected, we instead set $\whS^{\text{ICP}}(\mE) = [m]$ in this case. 
This choice aligns with the mathematical convention that the intersection over an empty index set is the universal set and is also more natural in interpretation: a variable is not regarded as a causal predictor if there exists a set $S$ for which $H_{0,S}(\mE)$ is true but does not contain it. Hence, if no such set $S$ exists, no variable can be excluded. 
Moreover, as shown in Section~\ref{sec:MultiFrameICP}, this convention leads to a more concise connection with the multiple testing formulation. See Appendix~\ref{app:SicpDef2} for further discussion.

% Another possibility is to define $\whS^{\text{ICP}}(\mE) = \emptyset$ if all $H_{ 0, S }(\mE)$ are rejected. This is reasonable from an interpretation perspective, as rejecting all $H_{ 0, S }(\mE)$ suggests that no set of variables is invariant across environments. Consequently, returning an empty set is a logical conclusion, as no variable can be identified as causal. However, this definition complicates the connection to the multiple testing formulation, hence we refrain from using it in this paper (see Appendix~\ref{app:SicpDef2} for more details).

\end{itemize}
The FWER guarantee of $\whS^{\text{ICP}}(\mE)$ holds at level $\alpha$ because
\begin{align*} %\label{FWER-ICP}
	P \left( \whS^{\text{ICP}}(\mE) \subseteq S^* \right) =
	P \left( \underset{S: H_{ 0, S }(\mE) \text{ is not rejected }} {\bigcap} S \subseteq S^* \right)
	\geq P( H_{ 0, S^* }(\mE) \text{ is not rejected } )
	\geq 1-\alpha.
\end{align*}

\section{A multiple testing formulation of ICP}\label{sec:MultiFrameICP}

The original formulation of ICP described in Section~\ref{sec:recap}, while elegant and concise, may not be the most suitable for addressing the two problems described in Section~\ref{sec:intro}. In this section, we cast the original ICP approach within a multiple testing framework. This reformulation enables us to leverage established techniques from multiple testing to obtain both FDR control and simultaneous true discovery bounds for ICP.

Specifically, instead of directly estimating $S(\mE)$ as in \eqref{Shat:ICP}, we consider testing the following hypotheses:
\begin{align} \label{def:direct-hypothesis-i}
	H^*_{0, i }:   i \not\in S^*, \quad i \in [m].
\end{align}
That is, $X_i$ is not a causal predictor of $Y$.
% Recall that $\mN^* = [m] \setminus S^*$ is the set of null causal predictors.

%The idea is to obtain valid p-values for these hypotheses, then apply false discovery rate control procedures, such as the Benjamini-Yekutieli procedure \citep{benjamini2001control}, or construct simultaneous true discovery bounds using these p-values. These approaches may lead to more discoveries compared to $\whS^{\text{ICP}}(\mE)$, because they aim to control a less conservative error rate than FWER.

To this end, based on the p-values $p_{S}$ generated by the first step of ICP,
we propose the following p-value for $H^*_{0, i }$:
\begin{align} \label{p-value-direct-approach-i}
	p^*_i = \max_{S \subseteq [m] \setminus \{i\}} p_{S}.
\end{align}
%Note that $p^*_i$ is constructed based solely on testing for invariance, and thus can be viewed as fully based on the ICP framework.
% p_{[m]} is not used for any i, this is similar to the original ICP case that the test result of H_{0,[m]} does not affect \hat S^{ICP}, no matter it is rejected or accepted due to taking union. But p_{[m]} will be useful for p^*_{\emptyset} when we consider intersection p^*_S later. So somehow this means we really need to consider testing intersection hypotheses to fully make use of the information from testing invariance in ICP.
All p-values in this paper depend on the environment set $\mE$, but we suppress this dependence for simplicity in notation. 
We take the p-values $p_S$ as given from the ICP approach and do not elaborate on their calculation. For more details, please refer to \cite{peters2016causal} and \cite{ heinze2018invariant}.

As shown in the following proposition, $p^*_i$ is a valid p-value.
\begin{proposition} \label{prop: valid-pi}
	%The p-value $p^*_i$ defined in \eqref{p-value-direct-approach-i} is valid for $H^*_{0, i }$ defined in \eqref{def:direct-hypothesis-i}. That is, 
	For any $c \in (0,1)$,
	$P_{H^*_{0, i }}(p^*_i \leq c) \leq c$.
\end{proposition}

One may wonder how the original ICP method connects to this multiple testing formulation, and we now establish this connection. 
Since the original ICP method controls the FWER, one might expect that applying a standard FWER control procedure, such as the Bonferroni correction or Holm’s procedure \citep{holm1979simple}, would yield a set equivalent to \(\whS^{\text{ICP}}(\mE)\). However, this is not the case.
Perhaps surprisingly, $\whS^{\text{ICP}}(\mE)$ is in fact equivalent to the set $\whS(\mE)$, which is obtained by directly comparing $p_i^*$ to the nominal level, without applying any multiple testing correction:
\begin{align} \label{whS:directCompare}
	\whS(\mE) = \{ i: p^*_i \leq \alpha \}.
\end{align}

\begin{proposition} \label{prop: equiICP}
	$\whS^{\text{ICP}}(\mE) = \whS(\mE)$.
\end{proposition}
As a direct result, $\whS(\mE)$ possesses the FWER control guarantee.
\begin{proposition} \label{prop: whSFWERcontrol}
	For $\alpha \in (0,1)$, 
	we have
	$ P \left( \whS(\mE) \subseteq S^* \right)
	\geq 1-\alpha.$
\end{proposition}

Lastly, the multiple testing formulation enables us to consider testing the intersection hypothesis defined below, which plays a key role in obtaining FDR control and simultaneous bounds for ICP:
\begin{align} \label{direct-hypothesis-S}
	H^*_{0, S}:   i \not\in S^* \text{ for all } i \in S
\end{align}
for each $S \subseteq [m]$.
 
To this end, we generalize the p-value \eqref{p-value-direct-approach-i} and propose the following p-value for testing $	H^*_{0, S}$:
\begin{align} \label{p-value-direct-approach-S}
	p^*_S = \max_{I \subseteq [m] \setminus S} p_{I}.
\end{align}
The validity of $p^*_S$ is shown in the following proposition.
\begin{proposition}\label{prop: valid-pS}
	For any $c \in (0,1)$,
	$ P_{H^*_{0, S}}(p^*_S \leq c) \leq c $.
\end{proposition}
%%%%%%%%%%%%%%%%%%%%%%%%%%%%%%%%%%%%%
% \section{Equip ICP with (simultaneous) FDR control}
\section{Bringing (simultaneous) FDR control to ICP}\label{sec:subsecICPFDR}

\subsection{Challenges in obtaining non-trivial FDR control via p-values}

Proposition~\ref{prop: whSFWERcontrol} shows that directly comparing $p^*_1, \dots, p^*_m$ to the nominal level $\alpha$ already controls the FWER, as if there were no multiple testing issue. This is somewhat unexpected, since one would usually need a multiple testing adjustment, such as the Bonferroni correction, to obtain FWER control. Since FWER control implies FDR control, the same direct comparison also controls the FDR. This complicates the usual strategy for gaining power by relaxing the error criterion from FWER to FDR. Any improvement would necessarily reject some $H^*_{0,i}$ with $p_i^* > \alpha$. While FDR control procedures exist that sometimes allow such rejections \citep[e.g.][]{benjamini2006adaptive, solari2017minimally}, such procedures never uniformly improve on unadjusted testing and are much less powerful in practice.

Given this, one may wonder whether the construction~\eqref{p-value-direct-approach-i} of $p^*_i$ based on $p_S$ is conservative. 
If that were the case, and if one could construct smaller valid p-values for $H^*_{0,i}$ from $p_S$, we might return to the usual situation where direct comparison to the nominal level no longer controls the FWER, and applying an FDR control procedure could yield more discoveries than an FWER control procedure.

To this end, we consider an ideal case where the optimal $p_S$ is available for testing $H_{0,S}(\mE)$.
Namely, $p_S \sim U(0,1)$ under a true null hypothesis and $p_S = 0$ under a false null hypothesis. This is also called the Dirac-uniform configuration \citep{finner2009false}.
Then, for an identifiable causal predictor $i \in S(\mE)$ (see \eqref{def:IdentifiableCausalPredictor}), 
we have $p^*_i = \max_{S \subseteq [m] \setminus \{i\}} p_{S} = 0$, since $H_{0,S}(\mE)$ is false for any $S$ that does not contain $i$. % by definition.  %, and so $p_S=0$.
That is, every identifiable causal predictor receives the smallest possible p-value under construction \eqref{p-value-direct-approach-i}.
As such, \eqref{p-value-direct-approach-i} is asymptotically optimal if $p_S$ is asymptotically optimal as the sample size tends to infinity. Therefore, \eqref{p-value-direct-approach-i} does not appear to be a conservative construction.
As a result, it is rather unclear how to obtain non-trivial FDR control for ICP based on p-values $p^*_S$ or $p_S$.

% Note that obtaining these $p_S$ values is a separate task from constructing a valid p-value for the hypothesis \eqref{def:direct-hypothesis-i} based on them.
% The p-values of non-identifiable causal predictors or non-causal predictors might be large, which makes sense.
% In general, when $p_S$ is not optimal, we cannot completely rule out the possibility of constructing a more powerful p-value for $H^*_{0,i}$. Nevertheless, the discussion above suggests that any such improvement is unlikely to be substantial enough to yield more discoveries after applying an FDR control procedure.

Recently, \cite{xu2025bringing} proposed a fundamental multiple testing principle called the e-Closure principle, which proceeds through e-values \citep[see, e.g.,][]{shafer2021testing, grunwald2024safe, ramdas2025hypothesis} and has been shown to bring uniform improvement over some existing FDR control methods.
In particular, the e-Closure principle involves intersection hypotheses \eqref{direct-hypothesis-S}, and thus gives hope to obtain non-trivial FDR control for ICP, whose results were obtained merely from testing single hypotheses \eqref{def:direct-hypothesis-i}.
Next, we briefly recap the e-Closure principle.

% Simul FDR control?

%%%%%%%%%%%%%%%%%%
\subsection{The e-Closure principle and p-to-e calibrator} \label{sec:e-ClosureIntro}

The e-Closure principle is a general framework for multiple testing based on e-values.
For a null hypothesis $H_i$, an e-value $e_i$ is a nonnegative random variable satisfying $\mathbb{E}_{P}[e_i] \le 1$ for all $P \in H_i$.
The e-Closure principle takes as input an e-collection $E=(e_S)_{S\subseteq [m]}$, where $e_S$ is an e-value for the intersection hypothesis $H_S$.
Given an error metric $\mathrm{ER}_f$ and a nominal level $\alpha\in(0,1)$, it returns a set of candidate discovery sets
\begin{align}\label{def:eclosureset}
	\mR^{\mathrm{ER}_f}_{\alpha}(E)
	=
	\left\{
	R\subseteq [m] : e_S \ge \frac{f_S(R)}{\alpha}
	\text{ for all } S\subseteq [m]
	\right\}.
\end{align}
In particular, FWER and FDR are special cases of error metrics with 
\begin{equation} \label{def:RfwerRfdr}
    \begin{aligned}
	&\mR^{\FWER}_{\alpha}(E) = \left\{ R  \subseteq [m]: e_S \geq \frac{\indi_{ \{ |R \cap S| \geq 1\} }}{\alpha} \text{ for all } S \subseteq [m] \right \} \\
	\qquad\text{and}\qquad
	&\mR^{\FDR}_{\alpha}(E) = \left\{ R  \subseteq [m]: e_S \geq \frac{1}{\alpha} \frac{ |R \cap S| }{\max\{ 1, |R|\}} \text{ for all } S \subseteq [m] \right \}.
\end{aligned}
\end{equation}

\citet{xu2025bringing} proved that a set $R$ is obtainable by a procedure controlling $\mathbb{E}[f_S(R)]$ at level $\alpha$ if and only if there exists an e-collection $E$ such that $R \in \mathcal{R}^{\mathrm{ER}_f}_{\alpha}(E)$. Therefore, to obtain FDR-controlled sets for ICP, we construct an e-collection from ICP and compute $\mathcal{R}^{\FDR}_{\alpha}(E)$. In fact, leveraging the e-Closure principle in ICP yields more than standard FDR control. It provides simultaneous FDR control, meaning that the FDR is controlled simultaneously for all sets in $\mathcal{R}^{\FDR}_{\alpha}(E)$. Consequently, any set selected from $\mathcal{R}^{\FDR}_{\alpha}(E)$, even in a data-dependent manner, retains the FDR control guarantee. 

% Consequently, every set $R\in \mR^{\FWER}_{\alpha}(\mE)$ controls the FWER at level $\alpha$, and every set $R\in \mR^{\FDR}_{\alpha}(\mE)$ controls the FDR at level $\alpha$.

Under ICP, however, we obtain p-values $p^*_S$ rather than e-values.
To apply the e-Closure principle, these p-values must therefore be transformed into e-values via a p-to-e calibrator \citep{shafer2011test}, that is, a decreasing function $f:[0,1]\to[0,\infty]$ satisfying $\int_0^1 f(x)\,dx \le 1$. Different p-to-e calibrators can lead to different power properties, so the choice of calibrator is important.

The first p-to-e calibrator we may consider is the all-or-nothing calibrator
\begin{equation*}
	f^{0/1}_\alpha(x)=\frac{\mathbbm{1}\{x\le \alpha\}}{\alpha},
\end{equation*}
which yields the e-collection $E^{0/1}=(e_S^{0/1})_{S\subseteq [m]}$, where $ e_S^{0/1}=f^{0/1}_\alpha(p_S^*)=\frac{\mathbbm{1}\{p_S^*\le \alpha\}}{\alpha} \in \{0, \frac{1}{\alpha}\}$.
We show that applying e-Closure to $E^{0/1}$ yields a collection $\mR^{\FWER}_{\alpha}(E^{0/1})$ whose largest set is exactly the original FWER-controlled discovery set of ICP.
However, this calibrator is not useful for FDR control, since it yields exactly the same candidate discovery sets as under FWER control. This is formalized in the following proposition.
\begin{proposition}\label{prop:01-equals-fwer}
	Let $ \whS^{\text{ICP}}(\mE) = \{ i: p^*_i \leq \alpha \}$ be the discovery set of ICP, then we have
	$\mR^{\FDR}_{\alpha}(E^{0/1})
	=
	\mR^{\FWER}_{\alpha}(E^{0/1})
	=\{R\subseteq [m]: R\subseteq \whS^{\text{ICP}}(\mE) \}$.
\end{proposition}

%%%%%%%%%%%%%%%%%%%%%%%

To obtain non-trivial FDR control, we therefore need a different p-to-e calibrator.
Motivated by the Su calibrator from \citet{xu2025bringing}, we consider
\begin{equation}\label{eq:su-calibrator}
	f^{\mathrm{Su}}_\alpha(x) =
	\frac{1}{\max\{ \rho_\alpha x, \alpha\} },
	\qquad x\in[0,1],
\end{equation}
where $\rho_\alpha = -W_{-1}(-\alpha/e)$ and $W_{-1}$ denotes the lower branch of the Lambert $W$ function.
For example, $\rho_\alpha$ is approximately $7.638$, $5.744$, and $4.890$ for $\alpha=0.01$, $0.05$, and $0.1$, respectively.
This leads to the e-collection 
\begin{align*}
	E^{\mathrm{Su}}=(e_S^{\mathrm{Su}})_{S\subseteq [m]}, 
	\quad \text{with} \quad e_S^{\mathrm{Su}} = f^{\mathrm{Su}}_\alpha(p_S^*) = \frac{1}{ \max\{ \rho_\alpha p_S^*, \alpha\} }.
\end{align*}

Unlike the all-or-nothing calibrator, we can already construct examples in which the Su calibrator yields nontrivial FDR-controlled discovery sets compared to the original ICP procedure.
% via e-Closure, whereas the original ICP procedure returns no discoveries.
For example, consider a special case where $p_S>\alpha$ for all $S$ with $|S|=m-1$, while $p_S=0$ for all other $S$. Then $p_i^*>\alpha$ for all $i$ and $p_S^*=0$ for all other $S$, so the original ICP procedure returns the empty discovery set. In contrast, since the threshold used in $\mR^{\FDR}_{\alpha}(E^{\mathrm{Su}})$ decreases with $m$, $\mR^{\FDR}_{\alpha}(E^{\mathrm{Su}})$ can contain nonempty sets when $m$ is sufficiently large. 
This constructed example suggests that combining e-Closure with a suitable calibrator can indeed yield nontrivial FDR control for ICP.
In fact, we show through simulations that this improvement is not merely theoretical, and that using $\mathcal{R}^{\mathrm{FDR}}_{\alpha}(E^{\mathrm{Su}})$ can indeed yield more discoveries than the original ICP procedure in some settings.

%%%%%%%%%%%%%%%%%%%%%%%%%%%%%%%%%%%%%%%%%%%%
\subsection{Adapted Su calibrators}\label{subsec:adasu}

For a given collection of p-values, when constructing FDR-controlled discovery sets via e-Closure, the e-values obtained from the Su calibrator can be improved. This is because the thresholds against which the e-values are compared take only finitely many values \citep[see also][]{wang2022false, xu2025bringing}.
Specifically, for a fixed nonempty set $S \subseteq [m]$, the possible thresholds are of the form $j/(\alpha r)$, where $1 \le j \le |S|$ and $j \le r \le m - |S| + j$.
Since only these finitely many values affect the e-Closure decision rule, an efficient p-to-e calibrator should map p-values directly to them. In particular, it should take the form of a step function, otherwise part of the unit integral budget is spent on intermediate e-values that play no role in the decision and is therefore wasted.
Motivated by this observation, we propose two adapted versions of the Su calibrator below.

\paragraph{Step-Su calibrator}
The first calibrator, denoted by Step-Su, is constructed from the Su calibrator and aims to uniformly improve it with respect to FDR control.
We first approximate the Su calibrator by a step function whose values are aligned with the attainable thresholds that exceed its minimum value, and then reallocate the remaining integral budget to enlarge the region on which the calibrator attains its largest value $1/\alpha$, thereby shifting the step function to the right.

\begin{figure}
	\centering
	% \captionsetup{width=.85\linewidth}
	\includegraphics[scale=0.5]{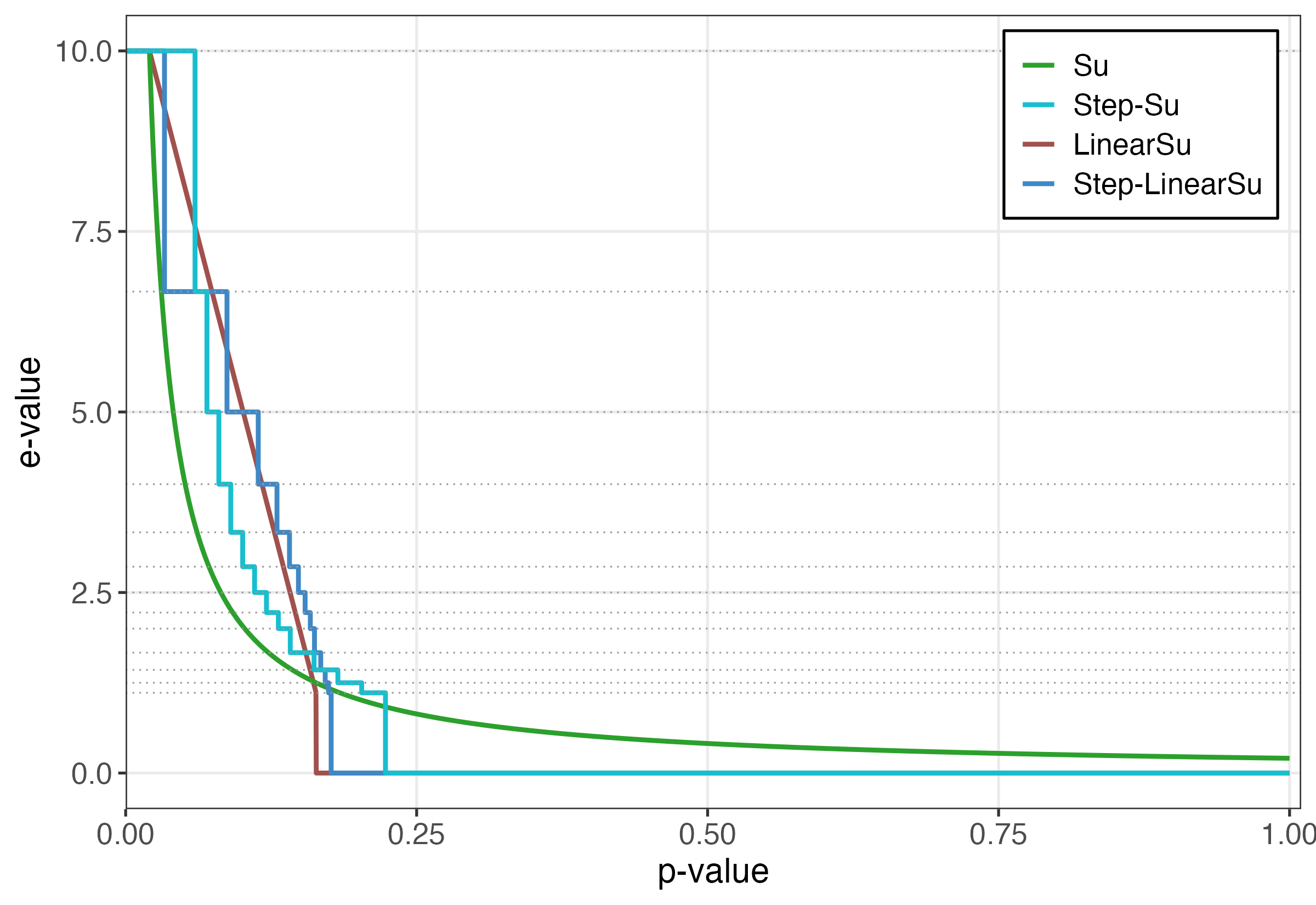}
	\caption[]
	{Comparison of the Su calibrator, Step-Su calibrator, LinearSu calibrator and Step-LinearSu calibrators for the case $m=10$, $|S|=2$, and $\alpha=0.1$. The dotted horizontal lines show the attainable e-Closure thresholds in $\mR^{\mathrm{FDR}}_{\alpha}(E)$ against which $e_S$ is compared.}
	\label{fig:calibrators}
\end{figure}

Consequently, the Step-Su calibrator is defined as
\[
f_{S,\alpha}^{\mathrm{Su,step}}(x)
=
\nu_{S,1}
\mathbf 1\{ 0 \le x\le y_{S,1}(a_S^*)\} + 
\sum_{\ell=2}^{L_S}\nu_{S,\ell}
\mathbf 1\{y_{S,\ell-1}(a_S^*)<x\le y_{S,\ell}(a_S^*)\},
%\sum_{\ell=1}^{L_S}\nu_{S,\ell} \mathbf 1\{y_{S,\ell-1}(a_S^*)<x\le y_{S,\ell}(a_S^*)\}.
\]
where $\{\nu_{S,\ell}\}_{\ell=1}^{L_S}$ are the attainable thresholds larger than the minimum value of the Su calibrator, and $L_S$ denotes their number. The values $\{y_{S,\ell}(a_S^*)\}_{\ell=1}^{L_S}$ define the jump locations of the step function, and $a_S^*$ is chosen such that the calibrator integrates to one.
See Appendix~\ref{app:stepsu} for more details.
A comparison between the Step-Su and Su calibrators is shown in Figure~\ref{fig:calibrators}.

Let $E^{\mathrm{Su,step}}$ and $E^{\mathrm{Su}}$ denote the e-collections obtained by applying $f_{S,\alpha}^{\mathrm{Su,step}}$ and $f_{\alpha}^{\mathrm{Su}}$ to the p-values $p^*_S$, respectively. The following proposition shows that the resulting e-Closure procedure is uniformly improved under FDR control.
\begin{proposition}\label{prop:sustep-uniform}
	$
	\mathcal R^{\mathrm{FDR}}_\alpha( E^{\mathrm{Su}})
	\subseteq
	\mathcal R^{\mathrm{FDR}}_\alpha(E^{\mathrm{Su,step}}).
	$
\end{proposition}

%%%%%%%%%%%%%%
\paragraph{Step-LinearSu calibrator}

The Su calibrator may decay too quickly (see Figure~\ref{fig:calibrators}). As a result, $f^{\mathrm{Su, step}}_{S,\alpha}(x)$ may allocate too much integral budget to very small p-values. %, which are mapped to the e-value $1/\alpha$.
In addition, since ICP requires testing all $2^m$ intersection hypotheses, it is typically applied in practice to a small number of variables selected after a screening step, leading to a small-$m$ setting.
Taking these considerations into account, we further introduce a step-function version of the linear Su calibrator, denoted by $f^{\mathrm{LinearSu, step}}_{S,\alpha}(x)$, constructed from its linear counterpart $f^{\mathrm{LinearSu}}_{S,\alpha}(x)$.
This linear variant decays more slowly and allocates more budget to moderately small p-values. The Step-LinearSu calibrator is constructed according to the same principle as Step-Su, and we defer the details to Appendix~\ref{app:steplinearsu}.
Figure~\ref{fig:calibrators} provides a comparison of these two calibrators with the previous ones.

Let $E^{\mathrm{LinearSu}}$ and $E^{\mathrm{LinearSu,step}}$ denote the e-collections obtained by applying $f^{\mathrm{LinearSu}}_{S,\alpha}(x)$ and $f^{\mathrm{LinearSu, step}}_{S,\alpha}(x)$ to the p-values $p^*_S$, respectively. 
The next proposition shows that, for sufficiently small $m$, the resulting e-Closure procedures are uniformly improved under FDR control compared to those based on $E^{\mathrm{Su}}$.
\begin{proposition}\label{prop:adasu-uniform}
	Let $\kappa_\alpha = \rho_\alpha-1+\sqrt{(\rho_\alpha-1)^2+1}$.
	If $m \leq \kappa_\alpha$, then 
	$
	\mathcal R^{\mathrm{FDR}}_\alpha( E^{\mathrm{Su}})
	\subseteq
	\mathcal R^{\mathrm{FDR}}_\alpha(E^{\mathrm{LinearSu}}).
	%\subseteq \mathcal R^{\mathrm{FDR}}_\alpha(E^{\mathrm{LinearSu, step}}).
	$
	Moreover, for all $m$,
	$
	\mathcal R^{\mathrm{FDR}}_\alpha(E^{\mathrm{LinearSu}})
	\subseteq
	\mathcal R^{\mathrm{FDR}}_\alpha(E^{\mathrm{LinearSu, step}}).
	$
\end{proposition}
For $\alpha = 0.01, 0.05$, and $0.1$, the corresponding cutoff values are $\kappa_{0.01} = 13.35$, $\kappa_{0.05} = 9.59$, and $\kappa_{0.1} = 7.91$, respectively. Thus, for example, when $\alpha = 0.05$, the e-Closure procedure based on $E^{\mathrm{LinearSu, step}}$ uniformly improves upon that based on $E^{\mathrm{Su}}$ whenever $m \le 9$.

%$f^{\mathrm{LinearSu, step}}_{S,\alpha}(x)$ uniformly improves Su calibrator whenever $m \le 9$. 

\begin{remark}
    A key aspect in constructing efficient p-to-e calibrators for FDR control via e-Closure is how the unit integral budget is allocated across the finitely many relevant threshold levels.
    Our proposed calibrators provide two concrete allocation schemes: starting from the Su calibrator and its linear variant, we first distribute the budget according to their crossing points with these threshold levels, and then assign the remaining budget to the largest level.
    While there is no universally optimal allocation strategy, as it depends on the underlying distribution of non-null p-values, it would be interesting to develop improved p-to-e calibrators in settings where additional information is available.
\end{remark}

%%%%%%%%%%%%%%%%%%%%%%%%%%%%%%%
Finally, given $\mathcal{R}^{\FDR}_{\alpha}(E)$, one can also obtain a FWER-controlled set: the union of all singleton sets in $\mathcal{R}^{\FDR}_{\alpha}(E)$ satisfies FWER control at level $\alpha$, and this guarantee holds simultaneously with the FDR control at level $\alpha$ for all sets in $\mathcal{R}^{\FDR}_{\alpha}(E)$ (see Theorem~37 in \cite{xu2025bringing}). See Section~\ref{sec:numericalSimu} for a concrete simulation example illustrating simultaneous control of FDR and FWER.

\begin{remark}
In the ICP setting, we can show that for some p-to-e calibrator, the union of all singleton sets coincides with the ICP discovery set at a smaller nominal level $l_\alpha \leq \alpha$. However, the simultaneous validity of the FWER control at level $l_\alpha$ and the FDR control at level $\alpha$ for the other sets no longer holds, reflecting the price that must be paid for obtaining simultaneous FWER and FDR control. See Appendix~\ref{app:FWERandeclosureFDRrelation} for more details.
\end{remark}

%In particular, by choosing $\alpha$ appropriately (but independently of the data), we can recover the original ICP result at a desired FWER level while obtaining additional information. For example, to obtain a FWER-controlled set at level $0.1$, one may take $\alpha = 0.3303$ and compute $\mathcal{R}^{\FDR}_{\alpha}(E^{f^{\mathrm{Su}}_\alpha})$. The union of its singleton sets then yields a FWER-controlled set at level $l_\alpha \approx 0.1$. At the same time, $\mathcal{R}^{\FDR}_{\alpha}(E^{f^{\mathrm{Su}}_\alpha})$ contains additional sets with simultaneous FDR control at level $\alpha$, which can be used if the FWER-controlled set is too small or empty. From this perspective, it is always more advantageous to compute $\mathcal{R}^{\FDR}_{\alpha}(E^{f^{\mathrm{Su}}_\alpha})$ than to use the original ICP procedure. [this is incorrect, simultaneity of FWER at l_\alpha and FDR at \alpha does not hold.]
\section{Bringing simultaneous true discovery bounds to ICP} \label{sec:SimulTDICP}
% \section{Equip ICP with simultaneous true discovery bound}

In addition to FDR control, simultaneous true discovery bounds are another useful error control guarantee that is less conservative than FWER. 
In particular, by reformulating ICP as the multiple testing problem \eqref{def:direct-hypothesis-i} and constructing valid p-values $p^*_S$, we can readily leverage the closed testing procedure to obtain simultaneous true discovery bounds \citep{marcus1976closed, goeman2011multiple}.
% (see Appendix~\ref{app:reviewClsoedtesting} for a brief recap).
We give a brief recap of the closed testing procedure.

\subsection{A brief recap of closed testing} \label{app:reviewClsoedtesting}
The closed testing procedure was originally introduced for controlling the familywise error rate \citep{marcus1976closed}. Later, \cite{goeman2011multiple} proposed simultaneous false discovery bounds based on it.
Below, we present the three main steps of the closed testing procedure.

Consider a multiple testing problem with hypotheses $H_1, \dots, H_m$. 
For any $I \subseteq [m]$,
let $H_I = \cap_{i\in I} H_{i}$ be an intersection hypothesis,
where $H_I$ is true if and only if $H_i$ is true for all $i \in I$.
The closed testing procedure proceeds as follows:
\begin{itemize}
	\item[1.] Obtain local test results:
	Locally test all intersection hypotheses $H_I $ for $I \subseteq [m]$ at level $\alpha \in (0,1)$.
	\item[2.] Implement the closed testing adjustment: $H_I$ is rejected by closed testing if and only if $H_J$ is locally rejected in Step 1 for all $J \supseteq I$. 
    \item[3.] Obtain simultaneous false discovery bound $u^{ct}(R)$: For any set $R \subseteq [m]$, let $u^{ct}(R)$ be the size of the largest subset of $R$ that is not rejected by closed testing in Step 2. If no such subset exists, we set $u^{ct}(R)=0$.
    % use this convention to match the convention of S^{ICP}=[m] when all p_S <= alpha and \tildet_\alpha(S^{ICP}) = |S^{ICP}| below.
    % also the convention that the maximum over an empty collection is 0.
\end{itemize}
Then, \cite{goeman2011multiple} proved that $u^{ct}(R)$ is a simultaneous false discovery bound. That is,
\begin{align*}
	\Prob( |R \setminus S^*| \leq u^{ct}(R), \forall R\subseteq [m]) \geq 1-\alpha,
\end{align*}
where $S^*$ denotes the set of false null hypotheses.
% where $\mN \subseteq [m]$ is the set of true null hypotheses.

Equivalently, $t^{ct}(R) = |R| - u^{ct}(R)$ is a simultaneous true discovery bound satisfying 
\begin{align*}
	\Prob( |R \cap S^*| \geq t^{ct}(R), \forall R\subseteq [m]) \geq 1-\alpha.
\end{align*}
% In particular, by the definition of $u^{ct}(R)$, we have that $t^{ct}(R)$ is the size of the smallest subset of $R$ whose complement in $R$ is not rejected by closed testing. % in Step 2.

%%%%%%%%%%%%%%%%%%%%%%%%%%%%%%%%%%%%%%%%%%%%%%%%%%%%%%%%
\subsection{Simultaneous true discovery bound for ICP}
A key observation here is that due to the monotonicity of $p^*_{S}$, there is no need to implement the computationally expensive closed testing adjustment step. 
To see this, if $p^*_{S} \leq \alpha$, we must have $p^*_{S'} \leq \alpha$ for all $S \subseteq S'$,
%indicating that if a hypothesis is locally rejected, it must be rejected by closed testing. On the other hand, by definition, if a hypothesis is rejected by closed testing, it must be locally rejected.Therefore, a hypothesis is rejected by closed testing if and only if it is locally rejected,
indicating that a hypothesis is rejected by closed testing if and only if it is locally rejected,
hence no closed testing adjustment is needed in this case.
This leads to the following simultaneous true discovery bound:
\begin{align} \label{td-bound-CT}
	%t_{\alpha}(R) = \max \{|I|: I \subseteq R \text{ and } p^*_I > \alpha \}.
 t_{\alpha}(R) = \min \{|R \setminus I|: I \subseteq R \text{ and } p^*_I > \alpha \},
\end{align}
with the convention that $t_{\alpha}(R) = |R|$ if the feasible set is empty.
% use this convention to match the convention of S^{ICP}=[m] when all p_S <= alpha and \tildet_\alpha(S^{ICP}) = |S^{ICP}| below.
In words, $t_{\alpha}(R)$ is the size of the smallest subset of $R$ whose corresponding complement in $R$ has a p-value larger than $\alpha$. It can be seen as a shortcut of the standard closed testing procedure, which simplifies the computation by not requiring the full implementation of all the steps in closed testing.

The theorem below formally shows that $t_{\alpha}(R)$ is a simultaneous true discovery bound.
\begin{theorem}\label{thm:SimulBound}
	Let $\alpha \in (0,1)$,
	then
	\begin{align} \label{simultaneous-TD-ICP}
		P( |R \cap S^*| \geq t_{\alpha}(R) \text{ for any } R \subseteq [m])
		\geq 1-\alpha.
	\end{align}
\end{theorem}
Based on the simultaneous guarantee~\eqref{simultaneous-TD-ICP}, 
users can freely check any set $R$ they regard as interesting, and calculate a high-probability lower bound $t_{\alpha}(R)$ on the true discoveries. 

The simultaneous true discovery bound encompasses both FWER control and the defining sets \citep{heinze2018invariant} as special cases. 
Specifically, by searching for the largest set $R$ whose $t_{\alpha}(R) = |R|$, we obtain a set with FWER control; by searching for all the smallest sets $R$ with $t_{\alpha}(R) = 1$, we obtain the defining sets. 
In particular, as the following proposition shows, $t_{\alpha}(R)$ recovers the FWER control guarantee of the original ICP procedure.
\begin{proposition}\label{prop:recoverICPfwer}
$ t_\alpha(\widehat S^{\mathrm{ICP}})= |\widehat S^{\mathrm{ICP}}|.$
\end{proposition}
% it is easy to check that for $\whS^{\text{ICP}}$, we have $t_{\alpha}(\whS^{\text{ICP}}) = |\whS^{\text{ICP}}|$, thereby recovering the FWER control guarantee of the original ICP procedure.
Hence, the same causal discoveries and information obtained from the original ICP approach are fully retained.
This indicates that using our proposed simultaneous true discovery bound to check more sets is always beneficial, which will only extract additional useful information from data. See Section~\ref{sec:SimuAndRealData} for illustrations.
% Additionally, we would like to mention that simultaneous false discovery bounds can be obtained from the defining sets of \cite{heinze2018invariant} through a method called interpolation \citep{goeman2021only, jointError}.

In fact, one can also obtain a simultaneous true discovery bound based on the original ICP formulation, without resorting to the multiple testing framework. %\footnote{We thank an anonymous referee for pointing this out.}
Specifically, let
\begin{align}\label{formula:DirectSimulBound}
	%\tildet_\alpha(R) = \underset{S \subseteq [m] \text{ with } p_S > \alpha}{\max} |R \setminus S|,
 \tildet_\alpha(R) = \underset{S \subseteq [m] \text{ with } p_S > \alpha}{\min} |R \cap S|,
	\quad \text{for all } R \subseteq [m],
\end{align}
with the convention that $\tildet_\alpha(R)=|R|$ if no such set $S$ exists. % use this convention to match the convention of S^{ICP}=[m] when all p_S <= alpha and \tildet_\alpha(S^{ICP}) = |S^{ICP}| below.
Then, we have
\[
P( |R \cap S^*| \geq \tildet_\alpha(R) \text{ for any } R \subseteq [m]) \geq P( H_{ 0, S^* }(\mE) \text{ is not rejected} ) \geq 1-\alpha,
\]
where the first inequality holds by the definition of $\tildet_\alpha(R)$.  % use convention $\tildet_\alpha(R)=|R|$ to match the convention of S^{ICP} when all p_S <= alpha.

Although obtained based on different ideas, $t_\alpha(R)$ and $\tildet_\alpha(R)$ (see~\eqref{td-bound-CT} and \eqref{formula:DirectSimulBound}) are actually equivalent, as shown in the following theorem.
\begin{theorem}\label{thm:EquiBounds}
	Let $\alpha \in (0,1)$.
	For any $R \subseteq [m]$,
	\begin{align*} 
		 t_\alpha(R) =\tildet_\alpha(R).
	\end{align*}
\end{theorem}

It is not uncommon in the literature that two seemingly different methods are actually equivalent.
For example, the simultaneous bound based on closed testing from \cite{goeman2011multiple} is equivalent to the bound from \cite{genovese2006exceedance}, which does not use closed testing \citep{goeman2021only}.
In particular, the simultaneous bounds in this section can also be derived using the e-Closure principle \citep{xu2025bringing}.
Recognizing such connections is valuable because it provides different perspectives on the same method and allows insights to be gained from different angles.
For example, in our case, understanding that $\tildet_\alpha(R)$ is equivalent to a closed testing bound helps clarify its admissibility \citep{goeman2021only}, which  is rather unclear from its original formulation~\eqref{formula:DirectSimulBound}.

\section{Simulations and a real data application}\label{sec:SimuAndRealData}

We implement simulations to verify the FDR and simultaneous guarantees of our proposed methods, examine their performance compared to the original ICP procedure, and illustrate the additional causal information one may obtain from these less conservative guarantees in a real application. 
All simulations were carried out in R, and the code is available at \url{https://github.com/Jinzhou-Li/ICP-FDR-SimulBounds}.

\subsection{Numerical simulations} \label{sec:numericalSimu}
We generate data from a linear structural equation model with multiple environments:
\[
X^e \leftarrow BX^e + \epsilon^e + b^e,
\]
where $e$ is the environment index, $X^e = (X^e_1, \dots, X^e_{11})^\top \in \mathbb{R}^{11}$, $B \in \mathbb{R}^{11 \times 11}$, 
$\epsilon^e \sim N_{11}(0, \diag(\sigma))$ with $\sigma_j \sim \mathrm{Unif}(0.5,1)$ independently, and $b^e \in \mathbb{R}^{11}$ denotes the environment-specific mean-shift intervention.
We take $X_9$ as the response variable and let $X_1, \ldots, X_8$ be its direct causes, so that there are eight non-null causal predictors and two null predictors.

We consider one observational environment with $b^e = 0$ and ten interventional environments, one for each non-response variable, yielding $11$ environments in total. 
In each interventional environment, exactly one non-response variable is subject to a mean shift of size $\delta$ (the intervention strength), while all other coordinates remain unchanged. That is, $b^e$ has a single non-zero entry equal to $\delta$ at a non-response variable. 
For each environment, we generate $n = 50$ samples.
We conduct simulations in low ($\delta = 1$), medium ($\delta = 1.5$), and high ($\delta = 2$) signal regimes, each with varying nominal levels $\alpha \in \{0.05, 0.1, 0.15, 0.2 \}$.

We implement $200$ replications. 
In each replication, we first generate the matrix $B$ as a strictly lower-triangular matrix. 
We set $B_{9,1:8}$ to be nonzero to encode the direct causal effects. 
For the remaining lower-triangular entries, each element is set to be nonzero independently with probability $s_B = 0.5$, and all nonzero entries are sampled from $\mathrm{Unif}(1,2)$. 
Finally, we permute the variables as $(X_4, X_{11}, X_9, X_2, X_7, X_8, X_{10}, X_3, X_6, X_1, X_5)$ so that they are not in a causal ordering, with $B$, $b^e$, and $\epsilon^e$ permuted accordingly.

For each generated dataset, we apply the original ICP method and four e-Closure procedures based on the Su, Step-Su, LinearSu, and Step-LinearSu calibrators, and report the number of true discoveries for ICP and for the largest discovery sets returned by the e-Closure procedures. We also report simultaneous true discovery lower bounds and false discovery upper bounds for all $2^{10} = 1024$ subsets using the same p-values $p_S$ as in ICP.

%%%%%%%%%%%%%%%%%%%%%%%
\paragraph{Simulation results on FDR control}
\begin{figure}[t]
	\centering
	\includegraphics[width=\textwidth]{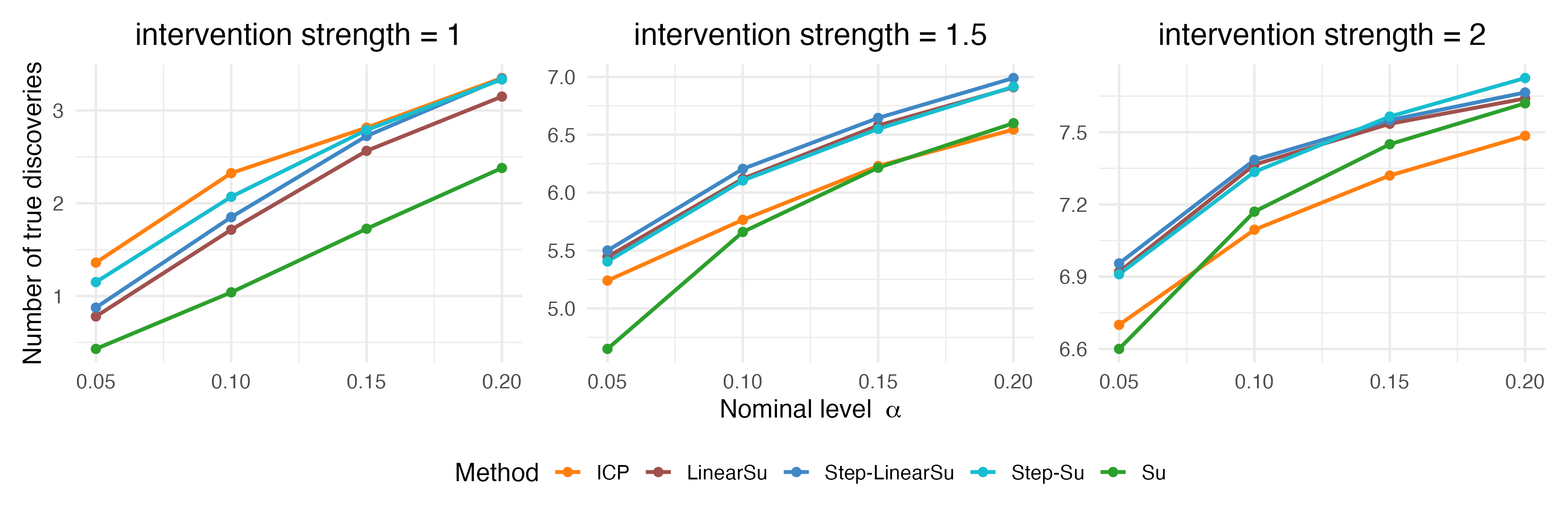}
	\caption{Average number of true discoveries for ICP and four e-Closure FDR methods with different p-to-e calibrators, averaged over 200 replications. Results are shown across different intervention strengths and varying nominal levels $\alpha$.}
	\label{fig:s05_TD_fdr}
\end{figure}

The empirical FDRs of all e-Closure-based FDR-controlling approaches are below the nominal level across all settings (see Appendix~\ref{appendix:simuError}), thereby validating the theoretical FDR control guarantee.
The average numbers of true discoveries for the original ICP procedure and the four e-Closure-based methods are shown in Figure~\ref{fig:s05_TD_fdr}. 

We observe that controlling FDR for ICP via e-Closure with suitable calibrators can indeed yield more true discoveries than the original ICP procedure in many settings. However, these FDR-controlling procedures may also yield fewer true discoveries than ICP, which controls FWER, especially in low-signal regimes with small nominal levels. This is not contradictory, as all these approaches can be formulated within the e-Closure framework with different e-collections and error metrics, and no general dominance in the number of discoveries exists when both differ.
% In addition, all methods yield more true discoveries as the nominal level and the intervention strength increase.

Among the four e-Closure approaches, the three methods (LinearSu, Step-Su and Step-LinearSu) whose calibrators explicitly account for the discrete thresholds used in the e-Closure for FDR control outperform the general-purpose Su calibrator which does not exploit this structure. Furthermore, the step-function calibrators (Step-Su and Step-LinearSu) consistently improve upon their non-stepwise counterparts (Su and LinearSu), indicating the benefit of a more efficient allocation of the integral budget.

%%%%%%%%%%%%%%%%%%%%%%%
\begin{figure}
\centering
    \includegraphics[width=0.78\textwidth]{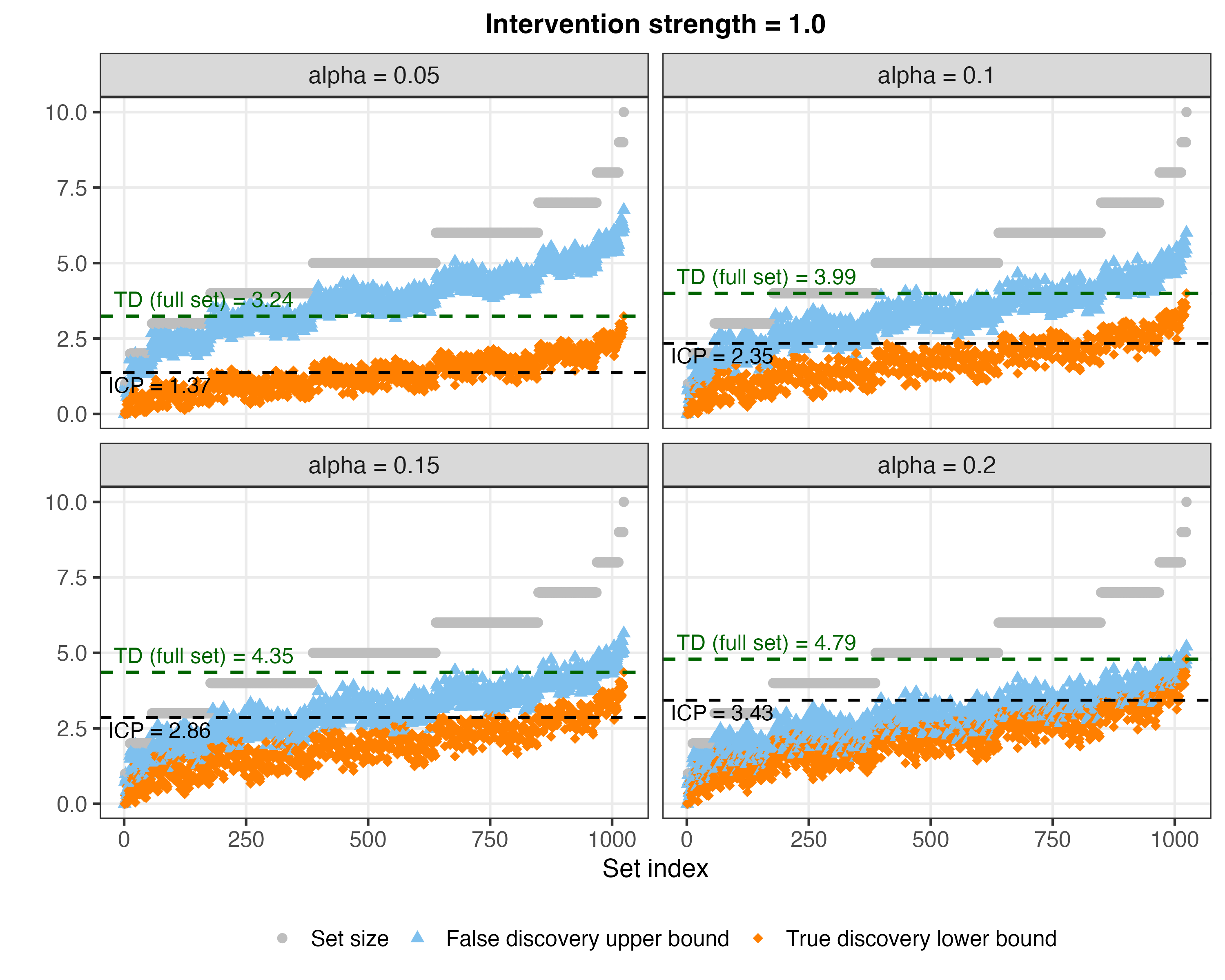}
\caption{Average simultaneous false discovery upper bounds and true discovery lower bounds over all $1024$ subsets, together with set sizes, under intervention strength $1$ at different nominal levels. The black dashed horizontal line indicates the average number of discoveries of the original ICP method, and the green dashed line indicates the average simultaneous true discovery lower bound for the full predictor set.}
    \label{fig:simulInt1}
\end{figure}

\begin{figure}[htbp]
\centering
\captionsetup[subfigure]{labelformat=simple,labelsep=space}
\renewcommand{\thesubfigure}{\Alph{subfigure}}
\begin{subfigure}[t]{0.49\textwidth}
    \centering
    \includegraphics[width=\textwidth]{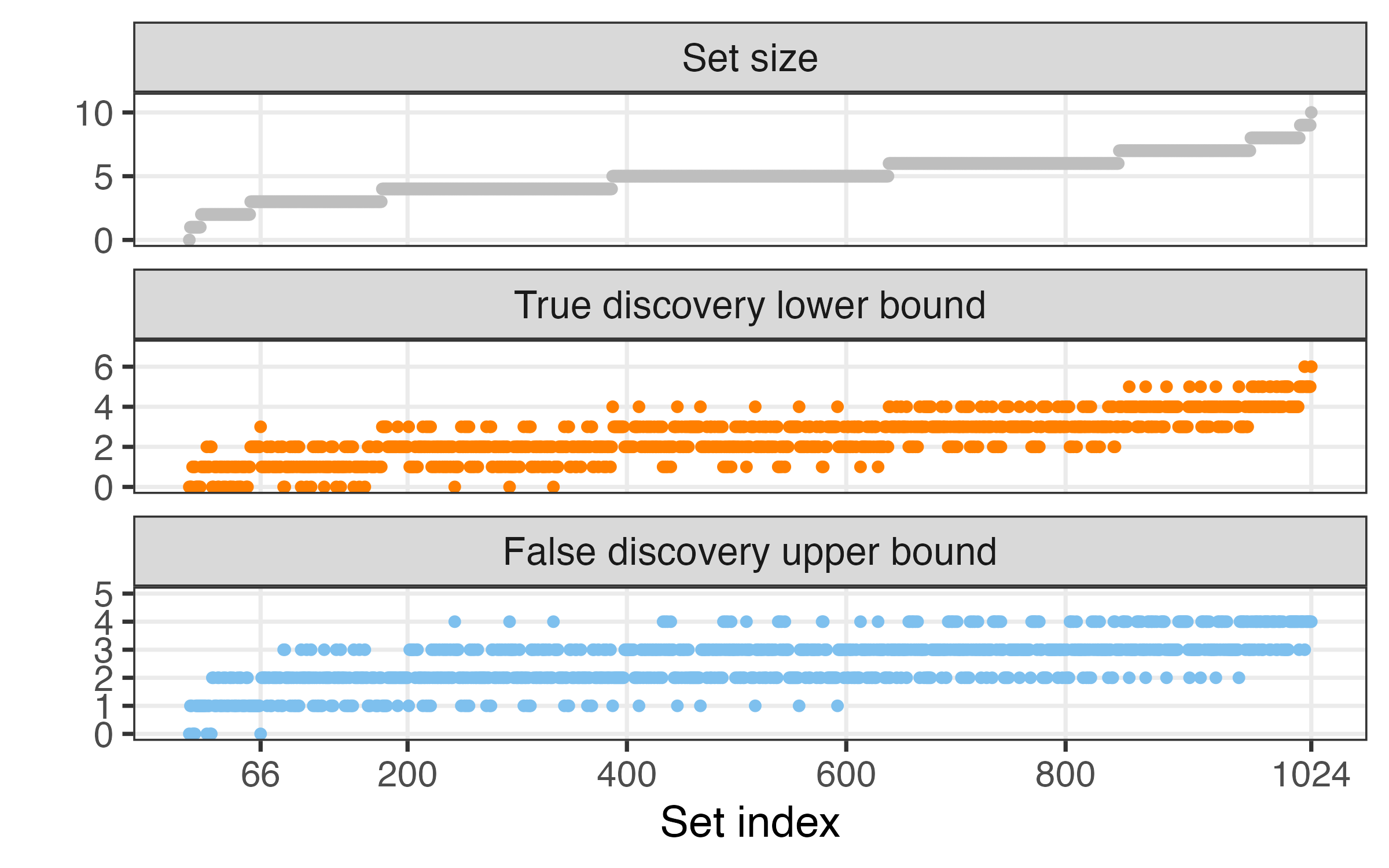}
    %\caption{$\alpha = 0.1$, $\delta = 1$}
    %\label{fig:simulOneData_A}
\end{subfigure}
\hfill
\begin{subfigure}[t]{0.49\textwidth}
    \centering
    \includegraphics[width=\textwidth]{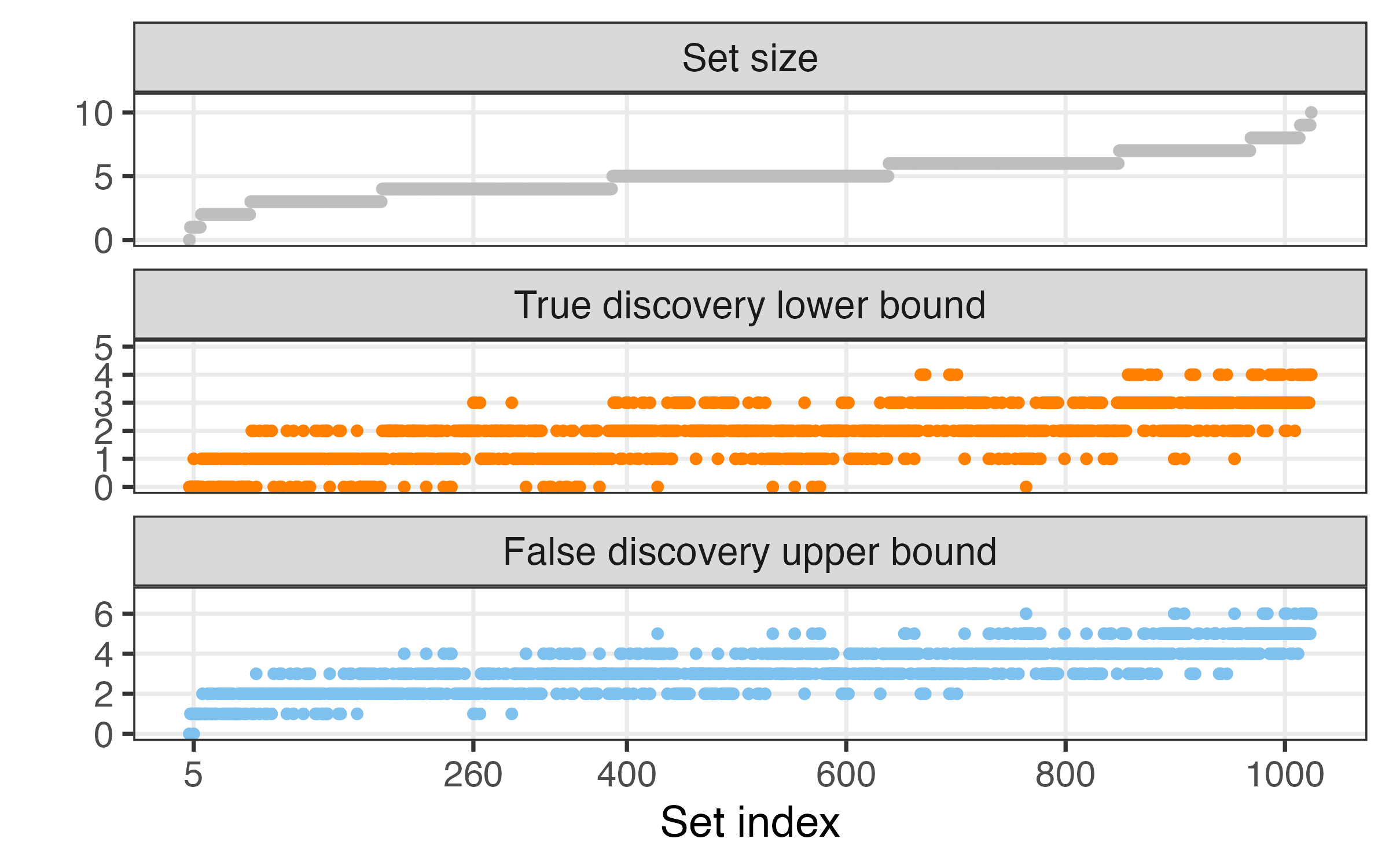}
    %\caption{$\alpha = 0.2$, $\delta = 1$}
    %\label{fig:simulOneData_B}
\end{subfigure}
\caption{Simultaneous false-discovery upper bounds, true-discovery lower bounds, and set sizes across all 1024 subsets for two simulated datasets. Panel A corresponds to a simulated dataset with $\alpha = 0.1$ and $\delta = 1$, while Panel B corresponds to a simulated dataset with $\alpha = 0.2$ and $\delta = 1$.}
\label{fig:simulOneData}
\end{figure}

%%%%%%%%%%%%%%%%%%%%%%%
\paragraph{Simulation results on simultaneous bounds}
As expected, the simultaneous guarantee~\eqref{simultaneous-TD-ICP} holds in all settings (see Appendix~\ref{appendix:simuError}). 
Figure~\ref{fig:simulInt1} shows the average simultaneous false discovery upper bounds and true discovery lower bounds across all $1024$ subsets, along with the corresponding set sizes, for the setting with intervention strength $1$. The simulation results for intervention strengths $1.5$ and $2$ are reported in Appendix~\ref{appendix:simuSimul}. 
The x-axis represents only the index of each set. For simplicity, subset compositions are omitted, as they do not affect the main point.
The black dashed horizontal line indicates the average number of discoveries of the original ICP method.

Compared to the original ICP method, the simultaneous bounds provide more insights into the underlying causal mechanism. For example, in the first plot with $\alpha=0.05$, the original ICP method yields $1.37$ discoveries on average, meaning that it identifies about $1.37$ causal predictors among the ten variables.
The simultaneous bounds also contain this information, since they assign a false discovery upper bound of zero to the set returned by the original ICP method in every replication. In addition, the simultaneous true discovery lower bound for the full predictor set (set index $1024$) shows that there are at least $3.24$ true causal predictors on average, thereby providing more informative causal insights.

%%%%%%%%%%%%%%%%%%%%
\paragraph{Illustration of the practical use of the new error criteria in two simulated datasets}

The above simulation results are averaged over $200$ replications. To better illustrate the practical use of the proposed (simultaneous) FDR control methods and simultaneous bounds, we look at two simulated datasets with true causal predictors (after permutation) $\{1,3,4,5,7,8,9,10\}$. For the e-Closure-based FDR control methods, we use the procedure with the Step-LinearSu p-to-e calibrator, which performs well across many settings in the simulation results presented before. 

The first simulated data set is generated in a setting with $\alpha = 0.1$ and $\delta = 1$.
In this case, the original ICP method returns the set $\{3,4,5\}$. Its FWER guarantee implies that, with probability at least $0.9$, all discovered variables are true causal predictors, which are indeed all true discoveries.
The Step-LinearSu method yields the following collection of candidate discovery sets:
\begin{align*}
&\{3\}, \{4\}, \{5\}, \{3,4\}, \{3,5\}, \{4,5\},
\{1,3,4\}, \{1,3,5\}, \{1,4,5\}, \{3,4,5\}, \\
&\{3,4,10\}, \{3,5,10\}, \{4,5,10\},
\{1,3,4,5\}, \{3,4,5,10\}, \{1,3,4,5,10\}.
\end{align*}
All these sets satisfy FDR control at level $0.1$ simultaneously, allowing researchers to select any of them for further study while retaining the same FDR guarantee. In particular, the largest set is $\{1,3,4,5,10\}$, for which the guarantee implies that the expected false discovery proportion is less than $0.1$. In this example, all five variables are true causal predictors, so ICP with FDR control identifies more true causal predictors than the original ICP method with FWER control. 
Moreover, as discussed in Section~\ref{subsec:adasu}, the union of all singleton sets among the candidate discovery sets satisfies FWER control at the same level. In this case, this set is $\{3,4,5\}$, which coincides with the output of the original ICP method, showing that the FDR-based approach recovers the original ICP result in this example. This recovery occurs coincidentally, as the FWER-controlled set obtained from the FDR-controlling method is generally smaller than the corresponding ICP discovery set at the same nominal level $\alpha$.

The simultaneous false discovery upper bounds and true discovery lower bounds for all $1024$ subsets are shown in Figure~\ref{fig:simulOneData}. 
The set with index $66$ corresponds to $\{3,4,5\}$, the discovery set returned by ICP. Its false discovery upper bound is zero, which recovers exactly the same guarantee as ICP.
Additional causal insights can be obtained by examining other sets. For example, the full set $\{1,2,3,4,5,6,7,8,9,10\}$ (set index $1024$) has a true discovery lower bound of $6$, meaning that, with probability at least $0.9$, at least six of the ten variables are true causal predictors. The recovery of the 0 false discovery upper bound for the ICP discovery set $\{3,4,5\}$ in this case is not coincidental, but guaranteed.

The second simulated data set is generated in a setting with $\alpha = 0.2$ and $\delta = 1$.
In this case, ICP returns a single discovery $\{4\}$. In contrast, Step-LinearSu returns $43$ candidate discovery sets, including sets of size seven such as $\{2,4,5,6,8,9,10\}$, which contains five true discoveries and two false discoveries. The simultaneous false discovery upper bounds and true discovery lower bounds are shown in Plot~B of Figure~\ref{fig:simulOneData}. 
As in the previous case, the simultaneous bounds recover the result of the original ICP procedure while providing additional information. For example, the true discovery lower bound for $\{4\}$ (with index 5) is one, while the true discovery lower bound for $\{1,4,6,9\}$ (with index 260) is three.
In particular, the corresponding $p_i^*$ values (see \eqref{p-value-direct-approach-i}) are $0.263, 0.246, 0.263, 0.049, 0.263$, $0.246, 0.263, 0.234, 0.223,$ and $0.263$. Hence, the additional causal information provided by both the FDR-controlling methods and the simultaneous bounds is not obtained simply by using a more liberal per-$p$-value threshold than the FWER-controlled ICP method. Rather, it arises from testing all intersection hypotheses $H^*_S$ (see \eqref{direct-hypothesis-S}) in both the FDR-controlling procedures and the construction of the simultaneous bounds, whereas the original ICP method with FWER control relies only on testing the individual hypotheses $H^*_i$ (see \eqref{def:direct-hypothesis-i}).

\subsection{Real data application}

We consider a real dataset used in \cite{stock2003introduction} about the educational attainment of teenagers in the US \citep{rouse1995democratization}.
This dataset contains information on $4739$ students from approximately 1100 US high schools, including gender, ethnicity, achievement test score, whether the father or mother is a college graduate, whether the family owns their home, whether the school is in an urban area, county unemployment rate, state hourly wage in manufacturing, average college tuition, whether the family income is above \$25000 per year, and region.

Following \cite{peters2016causal}, we consider two datasets from two environments, with $2231$ and $2508$ observations, respectively.
The response variable $Y$ is binary, indicating whether a student attained a BA degree or higher.
Categorical covariates are encoded using dummy variables, resulting in $13$ predictors in total.
We set the significance level to $\alpha = 0.1$.
Our goal is to identify the causal predictors of $Y$.

\begin{figure}
	\centering
	\includegraphics[width=0.8\linewidth]{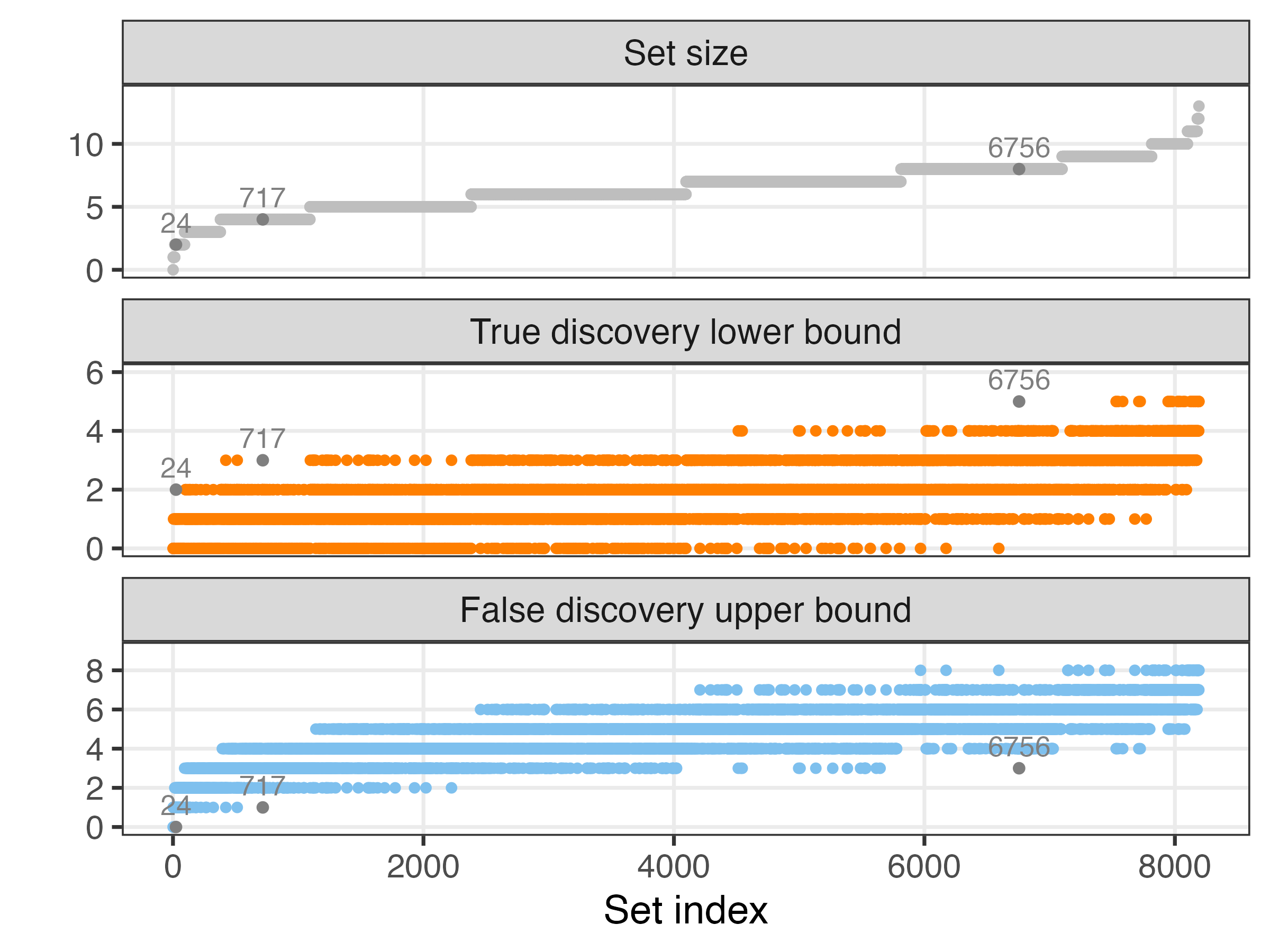}
	\caption{Simultaneous false discovery upper bounds, true discovery lower bounds, and set sizes over all $8192$ subsets based on the real data with $\alpha = 0.1$.}
	\label{fig:real}
\end{figure}

% We first apply ICP with a significance level $\alpha=0.05$, returning one variable `score' as the causal predictor. It is reasonable that test score has a causal influence on whether the student would attain a BA degree or higher.

% We then calculate the simultaneous bounds for all $2^{13}=8192$ sets, and the results are shown in Figure~\ref{fig:real}. More information can be obtained by looking at these bounds. For example, by looking at the set with index $112$, we know that with a probability larger than $0.95$, there are at least two causal predictors in \{`score', 'fcollege$\_$no', 'mcollege$\_$no'\}. It seems plausible that the true discovery lower bound is $2$ for this set, because the two variables of whether father or mother went to college are highly correlated, so it is difficult to distinguish between whether both are causal predictors or only one of them is.

We apply the original ICP method, which returns the set 
\{`score', `fcollege\_no'\} while controlling the FWER, as reported in \cite{peters2016causal}.
This implies that, with probability at least $0.9$, both variables are causal predictors of the response. This is reasonable, as a student’s test score and whether their father attended college are both likely to have a direct causal effect on whether the student attains a BA degree or higher.

We next apply e-Closure with the Step-LinearSu p-to-e calibrator to obtain FDR-controlled discovery sets. This yields two candidate sets \{`score', `fcollege\_no'\} and \{`ethnicity\_other', `score', `fcollege\_no'\}. The first coincides with the output of the original ICP method, while the second includes an additional binary variable, `ethnicity\_other', where `ethnicity\_other = 1' indicates that the student is neither African American nor Hispanic/Latino.
Interpreting `ethnicity\_other' as a causal predictor of whether the student attains a BA degree or higher is less plausible. Rather, this variable is likely acting as a proxy for other socio-economic factors, such as access to resources, school quality, or family background. Note that FDR control guarantees that the expected proportion of false discoveries among all discoveries is less than $\alpha$, and thus false discoveries may still be present in the selected set.
In addition, since there are no singleton sets among the candidate sets, no FWER-controlled set can be obtained from the FDR-controlling method in this case.

Finally, we compute simultaneous false discovery upper bounds and true discovery lower bounds for all $8192$ subsets, as shown in Figure~\ref{fig:real}. 
These simultaneous bounds contain the same information provided by the original ICP approach. In particular, the largest set whose false discovery upper bound is zero is \{`score', `fcollege\_no'\} (set index $24$), which coincides with the ICP discovery set and retains the same FWER guarantee.
Beyond this, the simultaneous bounds provide additional insight. For example, from Figure~\ref{fig:real}, the maximum true discovery lower bound across all subsets is five. Among the sets attaining this bound, the smallest one is \{`ethnicity\_other', `score', `fcollege\_no', `mcollege\_no', `urban\_no', `wage', `income\_low', `region\_other'\} (set index $6756$), which implies that at least five of these eight variables are true causal predictors with probability at least $0.9$.
Similarly, one can consider sets with a false discovery upper bound equal to one. One such set is \{`ethnicity\_other', `score', `fcollege\_no', `income\_low'\} (set index $717$), which implies that at most one variable in this set is not a causal predictor.
More information can be extracted from the simultaneous bounds, and the error control guarantee holds simultaneously over all sets.

\section{Discussion} \label{sec:discuss}

Motivated by the conservativeness of the FWER guarantee in the original ICP approach, we study how ICP can be equipped with less conservative error control guarantees, including FDR control and simultaneous true discovery bounds. Our key step is to reformulate ICP as a multiple testing problem for the null causal predictors, for which we propose valid p-values.
For FDR control, we show that directly comparing these p-values to the nominal level recovers the original ICP procedure, and hence controls both FWER and FDR, making non-trivial improvements based on p-values alone difficult. 
To obtain non-trivial FDR-controlling sets, we instead apply the e-Closure principle and develop new p-to-e calibrators tailored to the discrete threshold structure of the FDR criterion.
In parallel, we derive simultaneous true discovery bounds for ICP via closed testing, and show that the closed testing adjustment is in fact unnecessary in this setting. This yields a strictly more informative alternative to the original ICP output.
%These bounds extend the output of the original ICP procedure and can provide additional causal insight, thereby offering a strictly more informative alternative. 
Through simulations and a real data application, we demonstrate the practical benefits of these two less conservative error control criteria for ICP.

From a practical perspective, when choosing the error control in ICP, the original FWER control should not be used, as simultaneous true discovery bounds can recover the FWER-controlled set of the original ICP procedure while providing additional insight, and can therefore serve as a replacement. When a probability-type guarantee is desired, simultaneous true discovery bounds should be used to determine the discovery set. 
In contrast, when a more exploratory analysis is desired and an expectation-type guarantee is acceptable, ICP with simultaneous FDR control can be used and may yield larger discovery sets. In this case, we recommend using the Step-Su or Step-LinearSu calibrators as default choices. 
With several simultaneous FDR-controlled sets, the method may, additionally and simultaneously, also give a FWER-controlled set, though that set may be smaller than the FWER-controlled set of the original ICP method.

%%%%%%%%%%%%%%%%%%%%%

The original ICP method has been generalized to various settings, including non-linear models \citep{heinze2018invariant}, sequential data \citep{pfister2019invariant}, transformation models \citep{kook2023model}, and settings with a single environment and no interventions \citep{bodik2025structural}.
All these methods provide (asymptotic) FWER-type error control guarantees. It would be interesting to extend our proposed methods to these settings, either directly or through suitable generalizations, to achieve less conservative error control.

The ICP approach requires testing $2^{m}$ hypotheses, which becomes computationally intractable for large $m$ (or large $|R|$). This challenge is also inherent in our proposed methods for FDR control and simultaneous bounds, but unlike in other applications, e-Closure does not risk imposing additional computational complexity, which is already exponential for ICP. One way to circumvent this issue is to first reduce the set of potential causal predictors via pre-screening methods such as Lasso \citep[see][for further discussion]{peters2016causal}. 
There is also a substantial body of work proposing optimization-based methods built on the invariance principle to improve computational scalability \citep[e.g.,][]{ghassami2017learning, rothenhausler2019causal, rothenhausler2021anchor, fan2024environment, wang2024causal, shen2025causality}.
It would be interesting to investigate whether such methods can be equipped with rigorous error control guarantees.

The multiple testing formulation~\eqref{def:direct-hypothesis-i} provides a convenient way for error-controlled causal discovery by combining different sources of evidence from data. In this paper, we construct p-values for $H^*_{0,i}: i \not\in S^*$ using only evidence from invariance testing across environments by ICP. However, other sources of evidence, such as conditional independence tests, may also provide evidence against $H^*_{0,i}$. P-values constructed by combining evidence from multiple principles can then be directly incorporated into our methods to obtain FDR-controlled sets and simultaneous discovery bounds. Developing principled methods for combining such sources of evidence would be an interesting direction for future work. 

Finally, our application of e-Closure to ICP showcases the potential of using e-values and the e-Closure principle to construct new procedures in settings where p-value-based approaches may be difficult to develop. We used this principle to construct an FDR-controlling procedure in a situation where unadjusted p-values already controlled the FWER, and showed that the resulting method could reject more hypotheses than the FWER-controlling method.

%When testing the null hypothesis $H^*_{0,i}$ (see~\eqref{def:direct-hypothesis-i}), p-values in the ICP framework are constructed by testing invariance across different environments. However, other sources of information, such as conditional independence tests, may also provide evidence against $H^*_{0,i}$. If p-values are constructed by combining information from multiple principles when testing $H^*_{0,i}$, they can be directly used to obtain FDR-controlled sets via e-Closure and simultaneous discovery bounds via closed testing. It is unclear how to achieve this within the original ICP formulation, which highlights another advantage of the multiple testing formulation~\eqref{def:direct-hypothesis-i}. Developing principled methods for combining multiple criteria to test $H^*_{0,i}$ while controlling errors in causal discovery is an interesting direction for future research.

% Finally, our application of e-Closure to ICP showcases the potential of the e-Closure principle. We used this principle to construct an FDR-controlling procedure in a situation in which unadjusted testing already controlled the FWER, and showed that the resulting method was able to reject more hypotheses than the FWER-controlling method.

% To avoid exhaustive combinatorial search, 

%%%%%%%%%%%
\section*{Acknowledgment}
Jinzhou Li thanks Nicolai Meinshausen for helpful discussions and suggestions on an earlier draft of this paper. 
The authors thank an anonymous referee for pointing out the simultaneous true discovery bound based on the original ICP formulation.
J.L. gratefully acknowledges support from SNSF Grant P500PT-210978 and the National University of Singapore Start-Up Grant A-0010272-00-00. J.L. would also like to thank the Isaac Newton Institute for Mathematical Sciences, Cambridge, for support and hospitality during the programme ``Causal inference: From theory to practice and back again'', where part of this work was undertaken.

%%%%%

\newpage
\bibliographystyle{apalike}
\bibliography{Reference}

\appendix
\newpage
\newpage
\appendix
\begin{center}
	{\large\bf SUPPLEMENTARY MATERIAL}
\end{center}
%%%%%%%%%%%%%%%%%%%%%%%%%%%%%%%%%%%%%%%%

\section{Equivalence under the empty-set convention and modified p-values} 
\label{app:SicpDef2}
In Section~\ref{sec:recap}, we use the convention that $\whS^{\text{ICP}}(\mE) = [m]$ when all $H_{0,S}(\mE)$ are rejected. If instead we follow \cite{peters2016causal} and set $\whS^{\text{ICP}}(\mE) = \emptyset$ in this case, a slightly modified p-value for hypothesis $H^*_{0, i }$ (see~\eqref{def:direct-hypothesis-i}) is needed to establish the equivalence with $\whS^{\text{ICP}}(\mE)$ and to handle this special case. Specifically, for a given $\tau \in (0,1)$, define
\begin{align}\label{pvalue2}
	\Tilde{p}^*_i(\tau) = \max \{ p^*_i, \Pi_{S \subseteq [m]} \mathbbm{1}_{ p_S \leq \tau} \}.
\end{align}
Note that $\Tilde{p}^*_i(\tau)$ is also a valid p-value by Proposition~\ref{prop: valid-pi} and the fact that $\Tilde{p}^*_i(\tau) \geq p^*_i$. 
The following proposition shows that, with this modification, $\whS^{\text{ICP}}(\mE)$, under the convention that $\whS^{\text{ICP}}(\mE) = \emptyset$ when all $H_{0,S}(\mE)$ are rejected, coincides with the set obtained by directly thresholding $\Tilde{p}^*_i(\alpha)$ at level $\alpha$. 
\begin{proposition} \label{prop: equiICP2}
	For $\alpha \in (0,1)$,
	%let $\Tilde{p}^*_i(\alpha)$ be defined in \eqref{pvalue2},
	let $\Tilde{S}(\mE) = \{ i: \Tilde{p}^*_i(\alpha) \leq \alpha \}$.
	%and $\whS^{\text{ICP}}(\mE)$ be defined in \eqref{Shat:ICP}, 
	Then $\Tilde{S}(\mE) = \whS^{\text{ICP}}(\mE)$.
\end{proposition}
As a result, $\Tilde{S}(\mE)$ inherits the FWER control guarantee.

% To avoid this additional complication when relating the original ICP approach to the multiple testing formulation, we adopt the convention $\whS^{\text{ICP}}(\mE) = [m]$ when all $H_{0,S}(\mE)$ are rejected.

%Note that $\whS^{\text{ICP}}(\mE) = \Tilde{S}(\mE) \subseteq \whS(\mE)$ as $p^*_i \leq \Tilde{p}^*_i(\alpha)$. Given that both $\whS(\mE)$ and $\whS^{\text{ICP}}(\mE)$ control the FWER, using $\whS(\mE)$ seems to improve upon the original ICP approach. However, $p^*_i < \Tilde{p}^*_i(\alpha) $ only occurs when $p_{S} \leq \alpha$ for all $S \subseteq [m]$. In such cases, $\whS^{\text{ICP}}(\mE) = \emptyset$ and $\whS(\mE) = [m]$, which seems useless because even $p_{S^*} \leq \alpha$, leading to a false rejection of $H_{0,S^*}(\mE)$. In practice, $\whS(\mE)$ and $\whS^{\text{ICP}}(\mE)$ are generally equivalent.

%%%%%%%%%%%%%%%%%%%%%%%%%%%%%%%%%%%%%%%%%%%%%%
\section{Details of the adapted Su calibrators}\label{app:adasu}

\subsection{Details of the Step-Su calibrator}\label{app:stepsu}
For a fixed nonempty set $S \subseteq [m]$, the possible thresholds $\frac{1}{\alpha} \frac{ |R \cap S| }{\max\{ 1, |R|\}}$ in e-Closure under FDR control are
\begin{align} \label{def:AllThreSet}
	\mathcal T_S = \left\{\frac{j}{\alpha r}: 1\le j\le |S|,\ j\le r\le m-|S|+j\right\}.
\end{align}
% which is attained by taking exactly one element from $S$ and all $m-|S|$ elements from $[m] \setminus S$ to form $R$. 

To ensure that the minimum value of the new (step-function) calibrator is no smaller than that of Su calibrator, let
\[
\lambda_S = \max \left(c_S,\frac{1}{\rho_\alpha}\right),
\]
where $ c_S = \frac{1}{\alpha(m-|S|+1)} \in \mathcal T_S$ is the smallest threshold, and $\frac{1}{\rho_\alpha}= f^{\mathrm{Su}}_\alpha(1)$ is the smallest value attained by Su calibrator.
Then, we define the image of the new calibrator as 
\[
\mathcal V_S
=
\{\tau\in \mathcal T_S:\tau>\lambda_S\}\cup\{\lambda_S\},
\]
and write the distinct elements of $\mathcal V_S$ in decreasing order as
\[
\frac{1}{\alpha}=\nu_{S,1}>\nu_{S,2}>\cdots>\nu_{S,L_S}=\lambda_S.
\]

For each $\ell = 1,\dots,L_S$, define the crossing point of $\nu_{S,\ell}$ with $f_\alpha^{\mathrm{Su}}$ as
\[
z_{S,\ell}
=
\sup\Bigl\{x \in [0,1] : f_\alpha^{\mathrm{Su}}(x) \ge \nu_{S,\ell}\Bigr\}
=
\frac{1}{\rho_\alpha \nu_{S,\ell}}.
%\quad \text{and set } z_{S,0} = 0.
\]

Based on these crossing points, we define the step function 
\[
g_{S,\alpha}^{\mathrm{Su,step}}(x)
=
\nu_{S,1} \mathbf 1\{0 \le x\le z_{S,1}\} +
\sum_{\ell=2}^{L_S}\nu_{S,\ell}
\mathbf 1\{z_{S,\ell-1}<x\le z_{S,\ell}\}.
\]

By construction, $ g_{S,\alpha}^{\mathrm{Su, step}}(x) = f_\alpha^{\mathrm{Su}}(x) = \alpha^{-1}$ on $[0,\frac{\alpha}{\rho_{\alpha}}]$ and $ g_{S,\alpha}^{\mathrm{Su, step}}(x) \leq f_\alpha^{\mathrm{Su}}(x)$ for all $x\in(\frac{\alpha}{\rho_{\alpha}},1]$.
The wasted integral budget of Su calibrator is 
\[
\int_0^1 f_\alpha^{\mathrm{Su}}(x)\ dx -\int_0^1 g_{S,\alpha}^{\mathrm{Su,step}}(x)\ dx = 
1-\int_0^1 g_{S,\alpha}^{\mathrm{Su,step}}(x)\ dx \geq 0.
\]

To make use of the wasted integral budget, we proceed by allocating it to the largest value $1/\alpha$. This enlarges the interval on which the new step-function calibrator equals $1/\alpha$, thereby shifting the remaining steps to the right. Consequently, the new calibrator is pointwise no smaller than $g_{S,\alpha}^{\mathrm{Su,step}}(x)$.
Formally, we define the new calibrator as
\[
f_{S,\alpha}^{\mathrm{Su, step}}(x; a) = 
\nu_{S,1}
\mathbf 1\{ 0 \le x\le y_{S,1}(a)\} + 
\sum_{\ell=2}^{L_S}\nu_{S,\ell}
\mathbf 1\{y_{S,\ell-1}(a)<x\le y_{S,\ell}(a)\},
\]
for some $a \in [0,1]$, where
\[
%y_{S,0}(a) = 0,
%\quad
y_{S,\ell}(a) = \min\{1, z_{S,\ell} + a\},
\quad \ell = 1,\dots,L_S.
\]
That is, the interval corresponding to the largest level $1/\alpha$ is enlarged by $a$, while all subsequent steps are shifted to the right by the same amount, with truncation at $x=1$ if necessary.

To determine $a$ such that $f_{S,\alpha}^{\mathrm{Su,step}}(x;a)$ has integral $1$, define
\[
A_S(a) =\int_0^1 f_{S,\alpha}^{\mathrm{Su,step}}(x;a)\,dx.
\]
Then $A_S(a)$ is continuous and nondecreasing in $a$, with
\[
A_S(0)=\int_0^1 g_{S,\alpha}^{\mathrm{Su,step}}(x)\,dx\le 1,
\quad \text{and} \quad
A_S(1)=\frac1\alpha>1.
\]
Hence there exists
\[
a_S^* =\inf\{a\in[0,1]:A_S(a)\ge 1\},
\]
and by continuity $A_S(a_S^*)=1$. 
Therefore, the final p-to-e calibrator is 
\[
f_{S,\alpha}^{\mathrm{Su,step}}(x)
=
f_{S,\alpha}^{\mathrm{Su,step}}(x;a_S^*)
=
\nu_{S,1}
\mathbf 1\{ 0 \le x\le y_{S,1}(a_S^*)\} + 
\sum_{\ell=2}^{L_S}\nu_{S,\ell}
\mathbf 1\{y_{S,\ell-1}(a_S^*)<x\le y_{S,\ell}(a_S^*)\}.
%\sum_{\ell=1}^{L_S}\nu_{S,\ell} \mathbf 1\{y_{S,\ell-1}(a_S^*)<x\le y_{S,\ell}(a_S^*)\}.
\]

%Based on the calibrator $f_{S,\alpha}^{\mathrm{Su,step}}(x)$, we obtain the e-collection
%\begin{align*}
%	E^{\mathrm{Su,step}}=(e_S^{\mathrm{Su,step}})_{S\subseteq [m]}, 
%	\quad \text{with} \quad e_S^{\mathrm{Su,step}} = f^{\mathrm{Su,step}}_{S,\alpha}(p_S^*).
%\end{align*}

%%%%%%%%%%%%%%%%%%%%%%%%%%%%%%%%%%%%%%%%%%%%%%%%%%%%%%%%%
\subsection{Details of the Step-LinearSu calibrator}\label{app:steplinearsu}

We define the linear variation of Su calibrator as
\begin{equation}\label{eq:adasu-calibrator}
	f^{\mathrm{LinearSu}}_{S,\alpha}(x)=
	\begin{cases}
		\alpha^{-1}, & 0\le x\le \beta_{1},\\
		\alpha^{-1}\left(1-\dfrac{1-\alpha c_S}{\beta_{2}-\beta_{1}}(x-\beta_{1})\right), & \beta_{1}<x\le \beta_{2},\\
		0, & x>\beta_{2},
	\end{cases}
\end{equation}
where $\beta_{1}=\alpha / \rho_\alpha$, $\beta_{2}=\frac{\alpha}{1+\alpha c_S}\left(2-\frac{1-\alpha c_S}{\rho_\alpha}\right)$ and $ c_S = \frac{1}{\alpha(m-|S|+1)} \in \mathcal T_S$ (see \eqref{def:AllThreSet}) is the smallest threshold.
By construction (see also Figure~\ref{fig:calibrators}), $f^{\mathrm{LinearSu}}_{S,\alpha}$ coincides with $f^{\mathrm{Su}}_\alpha$ on the plateau $[0,\beta_{1}]$, then decreases linearly to the level $c_S$, and is zero for $x>\beta_{2}$. 
Moreover, it is straightforward to verify that $\int_0^1 f^{\mathrm{LinearSu}}_{S,\alpha}(x)\,dx \le 1$, and hence $f^{\mathrm{LinearSu}}_{S,\alpha}$ is a valid p-to-e calibrator.

Then, based on $f^{\mathrm{LinearSu}}_{S,\alpha}$, we construct a step-function calibrator similarly as before.
Specifically, we take its image to be $\mathcal T_S$, the set of all thresholds used in e-Closure, and write the distinct elements of $\mathcal T_S$ in decreasing order as
\[
\frac{1}{\alpha} = \tau_{S,1} > \tau_{S,2} > \cdots > \tau_{S,K_S} = c_S.
\]

For $k = 1,\dots,K_S$, define the crossing points
\[
x_{S,k}
=
\sup\Bigl\{x \in [0,1] : f^{\mathrm{LinearSu}}_{S,\alpha}(x) \ge \tau_{S,k}\Bigr\}.
% \quad \text{and set } x_{S,0} = 0.
\]
Then, for $a\in[0,1]$, define the step function
\[
f^{\mathrm{LinearSu, step}}_{S,\alpha}(x;a)
=
\tau_{S,1}
\mathbf 1\{0 \le x\le y_{S,1}(a)\} + 
\sum_{k=2}^{K_S}\tau_{S,k}
\mathbf 1\{y_{S,k-1}(a)<x\le y_{S,k}(a)\},
\]
% where $ y_{S,0}(a)=0$ and 
where $y_{S,k}(a)=\min\{1,x_{S,k}+a\}$ for $k=1,\dots, K_S$.

Finally, define 
\[
f^{\mathrm{LinearSu, step}}_{S,\alpha}(x)
=
f^{\mathrm{LinearSu, step}}_{S,\alpha}(x;a_S^*)
= \tau_{S,1}
\mathbf 1\{0 \le x\le y_{S,1}(a_S^*)\} + 
\sum_{k=2}^{K_S}\tau_{S,k}
\mathbf 1\{y_{S,k-1}(a_S^*)<x\le y_{S,k}(a_S^*)\},
\]
where 
\[
a_S^*=\inf\{a\in[0,1]:A_S(a)\ge 1\}
\quad \text{with} \quad
A_S(a)=\int_0^1 f^{\mathrm{LinearSu, step}}_{S,\alpha}(x;a)\,dx.
\]
Note that $a_S^*$ exists and satisfies $A_S(a^*)=1$ since
$A_S(a)$ is continuous and nondecreasing, with $A_S(0)\le 1$ and $A_S(1) = 1/\alpha>1$. 
% See Figure~\ref{fig:calibrators} for a comparison of $f^{\mathrm{LinearSu, step}}_{S,\alpha}$ with other calibrators.

%%%%%%%%%%%%%%%%%%%%%%%%%%%%%%%%%%%%%%%%
\subsection{Relationship between the union of singleton sets and the original ICP discovery set} \label{app:FWERandeclosureFDRrelation}

As discussed in Section~\ref{subsec:adasu}, the monotonicity of $p^*_S$ in the ICP setting allows us to show that the union of all singleton sets coincides with the ICP discovery set at a smaller nominal level $l_\alpha \leq \alpha$, as stated in the following proposition.
\begin{proposition} \label{prop:FWERfromFDReclosure}
    Let $f_\alpha(x)$ be a p-to-e calibrator and define $l_{\alpha} = \max \{t \in [0,1]: f_\alpha(t) \geq 1/\alpha\}$. 
    Let $\mathcal{R}^{\FDR}_{\alpha}(E^{f_\alpha})$ be obtained by applying e-Closure to the e-collection $E^{f_\alpha}$ with $e^*_S = f(p^*_S)$ for $S \subseteq [m]$, where $p^*_S$ are p-values obtained from ICP. Then, the union of all singleton sets in $\mathcal{R}^{\FDR}_{\alpha}(E^{f_\alpha})$ coincides with the ICP discovery set at nominal level $l_\alpha$:
    \[
        \{i : p_i^* \leq l_\alpha\}
        =
        \bigcup_{i \in [m]}\bigl\{\{i\} : \{i\} \in \mathcal{R}^{\FDR}_{\alpha}(E^{f_\alpha})\bigr\}.
    \]
\end{proposition}

For the Su calibrator $f^{\mathrm{Su}}_\alpha(x)$, we have $l_\alpha = \alpha/\rho_\alpha \leq \alpha$ since $\rho_\alpha \geq 1$.  %  and $\rho_\alpha \to 1$ as $\alpha \to 1$
Hence, the union of singleton sets in $\mathcal{R}^{\FDR}_{\alpha}(E^{f^{\mathrm{Su}}_\alpha})$ satisfies FWER control at level $l_\alpha \leq \alpha$. 
However, as noted in the remark of Section~\ref{subsec:adasu}, although this FWER-controlled set coincides with the ICP discovery set at level $l_\alpha$, the simultaneous validity with the other FDR-controlled sets at level $\alpha$ no longer holds. Specifically, if the FDR-controlled sets are observed before deciding to select the union of singleton sets, then the valid FWER guarantee for this set is only at level $\alpha$, rather than $l_\alpha$. This reflects the price that must be paid for obtaining FWER control simultaneously with the FDR control at level $\alpha$.

For the other three p-to-e calibrators, which depend on the size of the set $S$ when transforming $p^*_S$ into an e-value, the equality in Proposition~\ref{prop:FWERfromFDReclosure} no longer holds. In particular, the union of the corresponding singleton sets contains, rather than coincides with, the original ICP discovery set at a certain nominal level, as shown in the proposition below.
% if want to be very explicit, we may write the p-to-e calibrator set not including S=\emptyset, but since for S=\emptyset, typically the constraint in e-Closure will be satisfied, so we may treat it as convention and do not discuss such case here for simplicity.
\begin{proposition}\label{prop:FWERfromFDReclosureAdaptedf}
    Let $\mathcal{F}_{\alpha} = \{f_{S,\alpha}\}_{S\subseteq [m]}$ be a set of p-to-e calibrators, $l_{S,\alpha} = \max \{t \in [0,1]: f_{S,\alpha}(t) \geq 1/\alpha\}$, and $L_{\alpha} = \min_{ S\subseteq [m]} l_{S,\alpha}$.
    Let $\mathcal{R}^{\FDR}_{\alpha}(E^{\mathcal{F}})$ be obtained by applying the e-Closure to the e-collection $E^{\mathcal{F}}$ with $e^*_S = f_{S,\alpha}(p^*_S)$ for $S \subseteq [m]$, where $p^*_S$ are p-values obtained from ICP. Then, the union of all singleton sets in $\mathcal{R}^{\FDR}_{\alpha}(E^{\mathcal{F}})$ contains the original ICP discovery set at nominal level $L_\alpha$:
    \[
    \{i : p_i^* \leq L_\alpha\}
    \subseteq
    \bigcup_{i \in [m]}\bigl\{\{i\} : \{i\} \in \mathcal{R}^{\FDR}_{\alpha}(E^{\mathcal{F}})\bigr\}.
    \]
\end{proposition}

%%%%%%%%%%%%%%%%%%%%%%%%%%%%%%%%%%%%%%%%
\section{Proofs}
\subsection{Proof of Proposition~\ref{prop: valid-pi}}

\begin{proof}(Proposition~\ref{prop: valid-pi})
	When $H^*_{0, i }$ is true, we have $S^* \subseteq [m]\setminus\{i\}$ as $i \not\in S^*$, so $p^*_i = \max_{S \subseteq [m] \setminus \{i\}} p_{S} \geq p_{S^*}$.
	Therefore, for any $c \in (0,1)$,
	\begin{align*}
		P_{H^*_{0, i }}(p^*_i \leq c) \leq P_{H^*_{0, i }}(p_{S^*} \leq c)
		\leq c,
	\end{align*}
	where the last inequality holds because $H_{ 0, S^* }(\mE)$ is true.
\end{proof}

%%%%%%%%%%%%%%%%%%%%%%%%%%%%%%%%%%%%%%%%
\subsection{Proof of Proposition~\ref{prop: equiICP} and~\ref{prop: equiICP2} }

We first introduce the following lemma, which gives the relation between $\whS(\mE) = \{ i: p^*_i \leq \alpha \}$ and $\whS^{\text{ICP}}(\mE)$ in cases where at least one $H_{0,R}(\mE)$ is not rejected. Note that this result is irrelevant to the definition of $\whS^{\text{ICP}}(\mE)$ in the case where all $H_{0,R}(\mE)$ are rejected.
\begin{lemma} \label{lemma: equiExceptOneCase}
	Assume that there exists $R \subseteq [m]$ such that $H_{0,R}(\mE)$ is not rejected.
	%, that is, $p_{R} > \alpha$. 
	Let $\whS(\mE) = \{ i: p^*_i \leq \alpha \}$. Then, $\whS(\mE) = \whS^{\text{ICP}}(\mE)$.
\end{lemma}

We give the proof of Lemma~\ref{lemma: equiExceptOneCase}.
\begin{proof}(Lemma~\ref{lemma: equiExceptOneCase})
Let
\[
\mathcal A_\alpha
=
\{S\subseteq [m]: H_{0,S}(\mE) \text{ is not rejected at level } \alpha\}
=
\{S\subseteq [m]: p_S>\alpha\}.
\]
By assumption, \(\mathcal A_\alpha\neq\emptyset\). Hence
\[
\widehat S_{\mathrm{ICP}}(\mE)
=
\bigcap_{S\in\mathcal A_\alpha} S .
\]
%Recall that
%\[
%\widehat S(\mE)=\{i\in[m]:p_i^*\le \alpha\},
%\qquad
%p_i^*
%=
%\max_{S\subseteq [m]\setminus\{i\}} p_S .
%\]
For any \(i\in[m]\), we have
\[
\begin{aligned}
i\in \widehat S(\mE)
&\iff p_i^*\le \alpha\\
&\iff \max_{S\subseteq [m]\setminus\{i\}} p_S\le \alpha\\
&\iff p_S\le \alpha \text{ for all } S\subseteq [m]\setminus\{i\}\\
% &\iff \text{there is no } S\in\mathcal A_\alpha \text{ such that } i\notin S\\
&\iff i\in S \text{ for all } S\in\mathcal A_\alpha\\
&\iff i\in \bigcap_{S\in\mathcal A_\alpha} S\\
&\iff i\in \widehat S_{\mathrm{ICP}}(\mE).
\end{aligned}
\]
Therefore, \(\widehat S(\mE)=\widehat S_{\mathrm{ICP}}(\mE)\).
\end{proof}

%%%%%%%%%%%%%%%%%%%%
We give the proof of Proposition~\ref{prop: equiICP}. Note that for this proposition we take $\whS^{\text{ICP}}(\mE) = [m]$ when all $H_{ 0, S }(\mE)$ are rejected.
\begin{proof}(Proposition~\ref{prop: equiICP})
	If there exists some $R \subseteq [m]$ such that $H_{0,R}(\mE)$ is not rejected, we have $\whS(\mE) = \whS^{\text{ICP}}(\mE)$ by Lemma~\ref{lemma: equiExceptOneCase}.
	
	Now we consider the remaining case that $H_{0,S}(\mE)$ is rejected for all $S \subseteq [m]$, that is, $p_S \leq \alpha$ for all $S \subseteq [m]$.
    We have $p^*_i \leq \alpha$ for all $i \in [m]$ by definition, which implies that $\whS(\mE) = [m] = \whS^{\text{ICP}}(\mE)$.
\end{proof}

%%%%%%%%%%%%%%%%%%%%
In the following, we prove Proposition~\ref{prop: equiICP2}. Note that for this proposition we take $\whS^{\text{ICP}}(\mE) = \emptyset$ when all $H_{ 0, S }(\mE)$ are rejected.
\begin{proof}(Proposition~\ref{prop: equiICP2})
	If there exists some $R \subseteq [m]$ such that $H_{0,R}(\mE)$ is not rejected, that is, $p_R > \alpha$.
	Then $\Tilde{p}^*_i(\alpha) = \max \{ p^*_i, \Pi_{S \subseteq [m]} \mathbbm{1}_{ p_S \leq \alpha} \} =
	\max \{ p^*_i, 0 \} = p^*_i$,
	so $\Tilde{S}(\mE) = \whS(\mE) = \whS^{\text{ICP}}(\mE)$, where the second equality is by Lemma~\ref{lemma: equiExceptOneCase}.
	
	Now we consider the remaining case that $H_{0,S}(\mE)$ is rejected for all $S \subseteq [m]$, that is, $p_S \leq \alpha$ for all $S \subseteq [m]$.
	Then, we have $\whS^{\text{ICP}}(\mE) = \emptyset$ by definition,
	and $\Tilde{p}^*_i(\alpha) 
	= \max \{ p^*_i, \Pi_{S \subseteq [m]} \mathbbm{1}_{ p_S \leq \alpha} \} 
	= \max \{ p^*_i, 1 \} = 1 > \alpha$ for any $i \in [m]$.
	Hence $\Tilde{S}(\mE) = \emptyset = \whS^{\text{ICP}}(\mE)$.
\end{proof}

%%%%%%%%%%%%%%%%%%%%%%%%%%%%%%%%%%%%%%%%

\subsection{Proof of Proposition~\ref{prop: whSFWERcontrol}}

\begin{proof}(Proposition~\ref{prop: whSFWERcontrol})
By Proposition~\ref{prop: equiICP}, we have 
\[
	P \left( \whS(\mE) \subseteq S^* \right)
    =P \left( \whS^{\text{ICP}}(\mE) \subseteq S^* \right)
	\geq 1-\alpha,
	\]
where the last inequality holds because $\whS^{\text{ICP}}(\mE)$ possesses FWER control.

\iffalse
	When $H_{0,S^*}(\mE)$ is not rejected, for any $i \not\in S^*$, we have
	$p^*_i = \max_{S \subseteq [m] \setminus \{i\}} p_{S} \geq p_{S^*} > \alpha$, so $i \not\in \whS(\mE)$.
	Hence, we have $ \whS(\mE) \subseteq S^* $.
	Therefore, 
	\[
	P \left( \whS(\mE) \subseteq S^* \right)
	\geq P \left( H_{0,S^*}(\mE) \text{ is not rejected} \right)
	\geq 1-\alpha.
	\]
\fi
\end{proof}

%%%%%%%%%%%%%%%%%%%%%%%%%%%%%%%%%%%%%%%%
\subsection{Proof of Proposition~\ref{prop: valid-pS}}

% The proof is similar to that of Proposition~\ref{prop: valid-pi}.
\begin{proof}(Proposition~\ref{prop: valid-pS})
	When $H^*_{0, S}$ is true, we have $S^* \subseteq [m]\setminus S$ as $S \cap S^* = \emptyset$, so $p^*_S = \max_{I \subseteq [m] \setminus S} p_{I} \geq p_{S^*}$.
	Therefore, for any $c \in (0,1)$,
	\begin{align*}
		P_{H^*_{0, S}}(p^*_S \leq c) \leq P_{H^*_{0, S}}(p_{S^*} \leq c)
		\leq c.
	\end{align*}
	% where the last inequality holds because $H_{ 0, S^* }(\mE)$ is true.
\end{proof}

%%%%%%%%%%%%%%%%%%%%%%%%%%%%%%%%%%%%%%%%
\subsection{Proof of Proposition~\ref{prop:01-equals-fwer}}
\begin{proof}(Proposition~\ref{prop:01-equals-fwer})
	For simplicity, we write $	e_S:=e_S^{0/1}$. Note that $e_S = f^{0/1}_{\alpha}(p_S^*) = \frac{\mathbbm{1}\{p_S^*\le \alpha\}}{\alpha}
	% = \frac{1}{\alpha}\mathbbm{1} \left\{ \max_{I\subseteq [m]\setminus S} p_I \le \alpha \right\}
	\in \left\{0, \frac{1}{\alpha} \right \}$ for any $S\subseteq [m]$.
	
	We first prove that $\mR^{\FDR}_{\alpha}(E^{0/1}) = \mR^{\FWER}_{\alpha}(E^{0/1})$.
	It suffices to show that for any fixed $R\subseteq [m]$ and any $S\subseteq [m]$, 
	\[
	e_S \geq \frac{\indi_{ \{ |R \cap S| \geq 1\} }}{\alpha} \Longleftrightarrow
	e_S \geq \frac{1}{\alpha} \frac{ |R \cap S| }{|R| \vee 1}.
	\]
	If $R\cap S = \emptyset$, then both of the above inequalities hold, as the right-hand sides are both zero.
	If $R\cap S \neq \emptyset$, the right-hand sides of the above two inequalities are both nonzero. Since $e_S \in \left\{0, \frac{1}{\alpha} \right\}$, the above two inequalities hold only when $e_S = \frac{1}{\alpha}$, and hence they are equivalent.
	
	Next, we prove that $\mR^{\FWER}_{\alpha}(E^{0/1}) =\{R\subseteq [m]: R\subseteq \whS^{\text{ICP}}(\mE) \}$.
	
	% If $\whS^{\text{ICP}}(\mE)= \{ i: p^*_i \leq \alpha \} = \emptyset$, then $\{R\subseteq [m]: R\subseteq \whS^{\text{ICP}}(\mE) \} = \{ \emptyset \}$ and $e_i=0$ for any $i \in [m]$. Then we must have $\mR^{\FWER}_{\alpha}(E^{0/1}) = \{ \emptyset \}$. Otherwise, if there exists some non-empty $S' \in \mR^{\FWER}_{\alpha}(E^{0/1})$, then taking $j \in S'$ and $R = \{j\}$ (see the definition of $\mR^{\FWER}_{\alpha}(E^{0/1})$), we have $0 = e_{j} \geq \frac{\indi_{ \{ |\{j\} \cap S'| \geq 1\} }}{\alpha} =  \frac{1}{\alpha}$, which is a contradiction. Now assume that $\whS^{\text{ICP}}(\mE) \neq  \emptyset$. [commented as no need to discuss this special case. The following works for $\whS^{\text{ICP}}(\mE)$ being both empty and non-empty]
	
	For any $R \in \mR^{\FWER}_{\alpha}(E^{0/1})$, we prove that $R\subseteq \whS^{\text{ICP}}(\mE)$.
	If $R = \emptyset$, it is clear that $R  \subseteq  \whS^{\text{ICP}}(\mE)$.
	If $R \neq \emptyset$, for any $i \in R$, we have 
	\[
	e_i
	\ge
	\frac{\mathbbm{1}\{|R\cap \{i\}|\ge 1\}}{\alpha}
	=
	\frac{1}{\alpha}.
	\]
	Then, because $e_i \in \left\{0, \frac{1}{\alpha} \right \}$, we have $e_i= \frac{1}{\alpha}$, which implies that $p^*_i \leq \alpha$ by definition.
	Hence $i \in \whS^{\text{ICP}}(\mE)$, and therefore $R \subseteq \whS^{\text{ICP}}(\mE)$.
	Thus, $\mR^{\FWER}_{\alpha}(E^{0/1}) \subseteq \{R\subseteq [m]: R\subseteq \whS^{\text{ICP}}(\mE)\}$.
	
	For any $R\subseteq \whS^{\text{ICP}}(\mE)$, we prove that $R \in \mR^{\FWER}_{\alpha}(E^{0/1})$.
	It suffices to show that for any $S\subseteq [m]$,
	\[
	e_S \ge \frac{\mathbbm{1}\{|R\cap S|\ge 1\}}{\alpha}.
	\]
	If $R\cap S=\emptyset$, then the right-hand side of the above inequality is $0$, so the inequality holds.
	If  $R\cap S\neq \emptyset$, take $i\in R\cap S$, so $i\in  \whS^{\text{ICP}}(\mE)\cap S$ as $R\subseteq \whS^{\text{ICP}}(\mE)$.
	Thus, $p_i^* \le \alpha$, and we have
	\[
	p_S^*
	=
	\max_{I\subseteq [m]\setminus S} p_I
	\le
	\max_{I\subseteq [m]\setminus\{i\}} p_I
	=
	p_i^* \le \alpha,
	\]
	which implies that  $e_S=1/\alpha$ by definition.
	Hence the inequality holds.
	
	Combining the above two cases, we conclude that $\mR^{\FWER}_{\alpha}(E^{0/1}) = \{R\subseteq [m]: R\subseteq \whS^{\text{ICP}}(\mE)\}$.
\end{proof}

%%%%%%%%%%%%%%%%%%%%%%%%%%%%%%%%%%%%%%%%
\subsection{Proof of Proposition~\ref{prop:sustep-uniform}}
\begin{proof}(Proposition~\ref{prop:sustep-uniform})
	For any $R\in \mathcal R^{\mathrm{FDR}}_\alpha(E^{\mathrm{Su}})$,
	it suffices to show that
	\[
	R\in \mathcal R^{\mathrm{FDR}}_\alpha(E^{\mathrm{Su,step}}).
	\]
	
	For any $S\subseteq [m]$, denote the threshold in e-Closure as
	\[
	\tau_S(R)=\frac{1}{\alpha}\frac{|R\cap S|}{|R|\vee 1}.
	\]
	For $S\subseteq [m]$ such that $R\cap S = \emptyset$, we have 
	\[
	e_S^{\mathrm{Su,step}} = f^{\mathrm{Su,step}}_{S,\alpha}(p_S^*)\ge \tau_S(R) = 0.
	\]
	Now consider $S\subseteq [m]$ such that $R\cap S\neq\emptyset$.
	
	If $\tau_S(R)\ge \lambda_S$, then $ \tau_S(R)=\nu_{S,\ell}$ for some $\ell \in \{1,\dots,L_S\}$.
	Since $R\in \mathcal R^{\mathrm{FDR}}_\alpha(E^{\mathrm{Su}})$, we have
	\[
	f_\alpha^{\mathrm{Su}}(p_S^*)\ge \tau_S(R)=\nu_{S,\ell},
	\]
    which implies
    \[
    p_S^*\le z_{S,\ell}
    \]
    by the definition of $z_{S,\ell}$.
    Since
    \[
    y_{S,\ell}(a_S^\star)=\min\{1,z_{S,\ell}+a_S^\star\}\ge z_{S,\ell},
    \]
    it follows that
    \[
    p_S^* \leq y_{S,\ell}(a_S^\star),
    \]
    and hence
    \[
    e_S^{\mathrm{Su,step}} =f^{\mathrm{Su,step}}_{S,\alpha}(p_S^*)\ge \nu_{S,\ell}=\tau_S(R)
    \]
    by the definition of $f^{\mathrm{Su,step}}_{S,\alpha}(x)$.
    %e_S^{\mathrm{Su,step}} =f^{\mathrm{Su,step}}_{S,\alpha}(p_S^*) \geq g_{S,\alpha}^{\mathrm{Su,step}}(p_S^*)\ge \nu_{S,\ell}
	
	If $\tau_S(R)<\lambda_S$, then $ \lambda_S=\frac{1}{\rho_\alpha}$,
	hence
	\[
	\tau_S(R)<\lambda_S=\frac{1}{\rho_\alpha} \leq f^{\mathrm{Su,step}}_{S,\alpha}(p_S^*) =e_S^{\mathrm{Su,step}}
	\]
	since the smallest positive value of $f^{\mathrm{Su,step}}_{S,\alpha}(x)$ on $[0,1]$ is $1/\rho_\alpha$.
	
	Combining the above cases, we have 
	\[
	e_S^{\mathrm{Su,step}} = f^{\mathrm{Su,step}}_{S,\alpha}(p_S^*)
	\ge
	\frac{1}{\alpha}\frac{|R\cap S|}{|R|\vee 1}
	\]
	for all $S\subseteq [m]$,
	so
	\[
	R\in \mathcal R^{\mathrm{FDR}}_\alpha(E^{\mathrm{Su,step}}).
	\]
\end{proof}

%%%%%%%%%%%%%%%%%%%%%%%%%%%%%%%%%%%%%%%%
\subsection{Proof of Proposition~\ref{prop:adasu-uniform}}

\begin{proof}(Proposition~\ref{prop:adasu-uniform})
	For 
	\[
	\beta_2 = \frac{\alpha}{1+\alpha c_S}\left(2-\frac{1-\alpha c_S}{\rho_\alpha}\right),
	\]
	we have
	\[
	f^{\mathrm{LinearSu}}_{S,\alpha}(\beta_2)=c_S,
	\]
	which is the smallest nonzero threshold against which $e_S$ is compared in e-Closure.
    % in $\mathcal R^{\mathrm{FDR}}_\alpha(E^{\mathrm{LinearSu}})$.
	Therefore, a sufficient condition for $\mathcal R^{\mathrm{FDR}}_\alpha( E^{\mathrm{Su}})
	\subseteq
	\mathcal R^{\mathrm{FDR}}_\alpha(E^{\mathrm{LinearSu}})$
	is that for every non-empty set $S\subseteq [m]$,
	\[
	f^{\mathrm{Su}}_\alpha(x)\le f^{\mathrm{LinearSu}}_{S,\alpha}(x)
	\quad \text{for all } x\le \beta_2.
	\]
	
	By construction, the two functions coincide on $[0,\beta_1]$. On $[\beta_1,\beta_2]$, the function $f^{\mathrm{LinearSu}}_{S,\alpha}$ is linear and $f^{\mathrm{Su}}_\alpha$ is convex, hence it suffices that
	\[
	f^{\mathrm{Su}}_\alpha(\beta_2)\le f^{\mathrm{LinearSu}}_{S,\alpha}(\beta_2),
	\]
	that is,
	\[
	\frac{1}{\rho_\alpha\beta_2}\le c_S.
	\]
	% for every non-empty set $S\subseteq [m]$.
	
	By calculation, this inequality is equivalent to
	\[
	m-|S|+1\le \kappa_\alpha,
	\]
	which is satisfied whenever
	\[
	m\le \kappa_\alpha.
	\]

    The second claim that $\mathcal R^{\mathrm{FDR}}_\alpha(E^{\mathrm{LinearSu}})
\subseteq
\mathcal R^{\mathrm{FDR}}_\alpha(E^{\mathrm{LinearSu, step}})$ can be proved analogously to Proposition~\ref{prop:sustep-uniform}, and is therefore omitted.

\end{proof}

%%%%%%%%%%%%%%%%%%%%%%%%%%%%%%%%%%%%%%%%
\subsection{Proof of Proposition~\ref{prop:FWERfromFDReclosure} and ~\ref{prop:FWERfromFDReclosureAdaptedf}}

The proofs of these two propositions follow similar arguments.
\begin{proof}(Proposition~\ref{prop:FWERfromFDReclosure})
By the definition of e-Closure, a singleton $\{i\}$ belongs to
$\mathcal{R}^{\FDR}_{\alpha}(E^{f_\alpha})$ if and only if
\[
    f_\alpha(p^*_S)
    =
    e^*_S
    \geq \frac{1}{\alpha}
    \qquad
    \text{for every } S \subseteq [m] \text{ with } i \in S,
\]
which is equivalent to
\begin{align} \label{proof:iInRFDRiff1}
    p^*_S \leq l_\alpha
    \qquad
    \text{for every } S \subseteq [m] \text{ with } i \in S
\end{align}
by the definition of $l_\alpha$.

For any $i$ such that $p_i^* \leq l_\alpha$, consider any set $S$ with $i \in S$. Then
\[
    p^*_S
    =
    \max_{I\subseteq [m]\setminus S} p_I
    \leq
    \max_{I\subseteq [m]\setminus\{i\}} p_I
    =
    p_i^* \leq l_\alpha,
\]
hence $\{i\}\in \mathcal{R}^{\FDR}_{\alpha}(E^{f_\alpha})$ by \eqref{proof:iInRFDRiff1}.

Conversely, if $\{i\}\in \mathcal{R}^{\FDR}_{\alpha}(E^{f_\alpha})$, then taking
$S=\{i\}$ gives $p_i^* \leq l_\alpha$. Therefore,
\[
    \bigcup_{i \in [m]}\bigl\{\{i\}: \{i\}\in \mathcal{R}^{\FDR}_{\alpha}(E^{f_\alpha})\bigr\}
    =
    \{i:p_i^*\leq l_\alpha\}.
\]
\end{proof}

\begin{proof}(Proposition~\ref{prop:FWERfromFDReclosureAdaptedf})
By the definition of e-Closure, a singleton $\{i\}$ belongs to
$\mathcal{R}^{\FDR}_{\alpha}(E^\mathcal{F})$ if and only if
\[
    f_{S,\alpha}(p^*_S)
    =
    e^*_S
    \geq \frac{1}{\alpha}
    \qquad
    \text{for every } S \subseteq [m] \text{ with } i \in S,
\]
which is equivalent to
\begin{align} \label{proof:iInRFDRiff}
    p^*_S \leq l_{S,\alpha}
    \qquad
    \text{for every } S \subseteq [m] \text{ with } i \in S
\end{align}
by the definition of $l_{S,\alpha}$.

For any $i$ such that $p_i^* \leq L_{\alpha} = \min_{S\subseteq [m]} l_{S,\alpha}$, consider any set $S$ with $i \in S$. Then
\[
    p^*_S
    =
    \max_{I\subseteq [m]\setminus S} p_I
    \leq
    \max_{I\subseteq [m]\setminus\{i\}} p_I
    =
    p_i^* \leq L_\alpha \leq l_{S,\alpha},
\]
hence $\{i\}\in \mathcal{R}^{\FDR}_{\alpha}(E^\mF)$ by \eqref{proof:iInRFDRiff}.
\end{proof}

%%%%%%%%%%%%%%%%%%%%%%%%%%%%%%%%%%%%%%%%
\subsection{Proof of Theorem~\ref{thm:SimulBound}}

This result directly follows from the fact that $t_{\alpha}(R)$ is equivalent to the simultaneous true discovery bound based on closed testing, as discussed in Section~\ref{sec:SimulTDICP}.
%So the simultaneous true discovery bound guarantee directly follows.
For completeness, we also provide a self-contained proof that does not rely on closed testing.
\begin{proof}(Theorem~\ref{thm:SimulBound})
    Let $\mN^*= [m] \setminus S^*$ be the set of null causal predictors.

    If $p^*_{\mN^*} > \alpha$, we have $|R \cap S^*| = |R \setminus \mN^*| \geq t_{\alpha}(R)$ by the definition of $t_{\alpha}(R)$, taking $I=R \cap \mN^*$.
 
	Hence
	\[
	P( |R \cap S^*| \geq t_{\alpha}(R) \text{ for any } R \subseteq [m])
	\geq 
	P( p^*_{\mN^*} > \alpha).
	\]
	In addition, we have 
	\begin{align*} 
		P( p^*_{\mN^*} \leq \alpha)
		= P( \max_{I \subseteq [m] \setminus \mN^*} p_{I} \leq \alpha) 
		\leq P( p_{S^*} \leq \alpha) 
		\leq \alpha,
	\end{align*}
	or equivalently
	\[
	P( p^*_{\mN^*} > \alpha) \geq 1-\alpha,
	\]
	which completes the proof.
\end{proof}

%%%%%%%%%%%%%%%%%%%%%%%%%%%%%%%%%%%%%%%%
\subsection{Proof of Proposition~\ref{prop:recoverICPfwer}}
\begin{proof}(Proposition~\ref{prop:recoverICPfwer})
Let
\[
A_\alpha=\{S\subseteq[m]:p_S>\alpha\}.
\]
If \(A_\alpha=\emptyset\), then \(\widehat S^{\mathrm{ICP}}=[m]\) and
\(p_U\le\alpha\) for all \(U\subseteq[m]\). Hence \(p_I^\ast\le\alpha\) for all
\(I\subseteq[m]\), so the feasible set in the definition of \(t_\alpha([m])\)
is empty. By convention,
\[
t_\alpha(\widehat S^{\mathrm{ICP}})
=
t_\alpha([m])
=
m
=
|\widehat S^{\mathrm{ICP}}|.
\]

Now suppose that \(A_\alpha\neq\emptyset\). Then
\[
\widehat S^{\mathrm{ICP}}=\bigcap_{S\in A_\alpha}S.
\]
Since \(A_\alpha\neq\emptyset\), the set \(I=\emptyset\) is feasible, because
\[
p_\emptyset^\ast=\max_{U\subseteq[m]}p_U>\alpha.
\]
Thus it remains to show that no nonempty
\(I\subseteq\widehat S^{\mathrm{ICP}}\) is feasible. Let
\(I\subseteq\widehat S^{\mathrm{ICP}}\) be nonempty and choose \(i\in I\).
For any \(U\subseteq[m]\setminus I\), we have \(i\notin U\). Since every set in
\(A_\alpha\) contains \(i\), such a set \(U\) cannot belong to \(A_\alpha\).
Therefore \(p_U\le\alpha\) for all \(U\subseteq[m]\setminus I\), and hence
\[
p_I^\ast=\max_{U\subseteq[m]\setminus I}p_U\le\alpha.
\]
So no nonempty \(I\subseteq\widehat S^{\mathrm{ICP}}\) is feasible. The only
feasible subset is \(I=\emptyset\), and consequently
\[
t_\alpha(\widehat S^{\mathrm{ICP}})
=
|\widehat S^{\mathrm{ICP}}\setminus\emptyset|
=
|\widehat S^{\mathrm{ICP}}|.
\]
\end{proof}

\begin{comment}
%%% Proof using Theorem 5.2
    \begin{proof}
Let
\[
A_\alpha=\{S\subseteq[m]:p_S>\alpha\}
\]
and define
\[
\widehat S^{\mathrm{ICP}}
=
\begin{cases}
\displaystyle\bigcap_{S\in A_\alpha}S, & A_\alpha\neq\emptyset,\\[6pt]
[m], & A_\alpha=\emptyset.
\end{cases}
\]
Then
\[
t_\alpha(\widehat S^{\mathrm{ICP}})
=
|\widehat S^{\mathrm{ICP}}|.
\]

By Theorem~5.2,
\[
t_\alpha(\widehat S^{\mathrm{ICP}})
=
\widetilde t_\alpha(\widehat S^{\mathrm{ICP}}).
\]
If \(A_\alpha=\emptyset\), then by convention
\[
\widetilde t_\alpha(\widehat S^{\mathrm{ICP}})
=
|\widehat S^{\mathrm{ICP}}|.
\]
If \(A_\alpha\neq\emptyset\), then
\[
\widehat S^{\mathrm{ICP}}\subseteq S
\qquad\text{for every }S\in A_\alpha.
\]
Therefore,
\[
|\widehat S^{\mathrm{ICP}}\cap S|
=
|\widehat S^{\mathrm{ICP}}|
\qquad\text{for every }S\in A_\alpha,
\]
and hence
\[
\widetilde t_\alpha(\widehat S^{\mathrm{ICP}})
=
\min_{S\in A_\alpha}|\widehat S^{\mathrm{ICP}}\cap S|
=
|\widehat S^{\mathrm{ICP}}|.
\]
Thus
\[
t_\alpha(\widehat S^{\mathrm{ICP}})
=
|\widehat S^{\mathrm{ICP}}|.
\]

\end{proof}
\end{comment}

%%%%%%%%%%%%%%%%%%%%%%%%%%%%%%%%%%%%%%%%
\subsection{Proof of Theorem~\ref{thm:EquiBounds}}

\begin{proof}(Theorem~\ref{thm:EquiBounds})
Let
\[
A(R)=\left\{I\subseteq R:\max_{U\subseteq[m]\setminus I}p_U>\alpha\right\}
\quad\text{and}\quad
B=\{S\subseteq[m]:p_S>\alpha\}.
\]
We first note that \(A(R)=\emptyset\) if and only if \(B=\emptyset\). To see this, if
\(B\neq\emptyset\), then
$ p_\emptyset^\ast=\max_{U\subseteq[m]}p_U>\alpha$,
so \(\emptyset\in A(R)\). Conversely, if \(A(R)\neq\emptyset\), then there exists
\(I\subseteq R\) such that $ \max_{U\subseteq[m]\setminus I}p_U>\alpha$,
and hence some \(U\subseteq[m]\setminus I\) satisfies \(p_U>\alpha\), implying
\(B\neq\emptyset\).

In the case that \(A(R)=\emptyset\), we have \(B=\emptyset\). By the conventions in the definitions
of \(t_\alpha(R)\) and \(\widetilde t_\alpha(R)\), we have
\[
t_\alpha(R)= \widetilde t_\alpha(R)=|R|,
\]
so the claim follows.

Now we consider the case that \(A(R)\neq\emptyset\), and hence \(B\neq\emptyset\).
Define
\[
u_\alpha(R)=|R|-t_\alpha(R)
= \max \{|I|: I \subseteq R \text{ and } p^*_I > \alpha \}
=\max_{I\in A(R)}|I|
\]
and
\[
\widetilde u_\alpha(R)=|R|-\widetilde t_\alpha(R)
= \max \{|R \setminus S|: S \subseteq [m] \text{ and } p_S > \alpha \}
=\max_{S\in B}|R\setminus S|.
\]
It suffices to show that \(u_\alpha(R)=\widetilde u_\alpha(R)\).

Let \(I_1\in A(R)\) and \(S_2\in B\) attain the two maxima, so that
\[
u_\alpha(R)=|I_1|,
\qquad
\widetilde u_\alpha(R)=|R\setminus S_2|.
\]
Set \(I_2=R\setminus S_2\). Since \(S_2\subseteq[m]\setminus I_2\) and
\(p_{S_2}>\alpha\), we have
\[
p_{I_2}^\ast
=
\max_{U\subseteq[m]\setminus I_2}p_U
\ge p_{S_2}
>
\alpha.
\]
Thus \(I_2\in A(R)\), and hence
\[
|R\setminus S_2|=|I_2|\le |I_1|.
\]

Conversely, since \(I_1\in A(R)\), there exists
\(S_1\subseteq[m]\setminus I_1\) such that \(p_{S_1}>\alpha\). Hence
\(S_1\in B\). Moreover, since \(I_1\subseteq R\) and \(S_1\cap I_1=\emptyset\),
we have \(I_1\subseteq R\setminus S_1\). Therefore,
\[
|I_1|
\le
|R\setminus S_1|
\le
|R\setminus S_2|,
\]
where the last inequality follows from the definition of \(S_2\). Combining the two
inequalities gives
\[
|I_1|=|R\setminus S_2|,
\]
and therefore
\[
u_\alpha(R)=\widetilde u_\alpha(R).
\]
%Equivalently,
%\[ t_\alpha(R)=\widetilde t_\alpha(R). \]
\end{proof}

%%%%%%%%%%%%%%%%%%%%%%%%%%%%%%%%%%%%%%%%
\section{Supplementary materials for simulations and real application} \label{appendix:simu}

\subsection{Simulation results on empirical error control performance} \label{appendix:simuError}

\begin{figure}
\centering
    \includegraphics[width=0.8\textwidth]{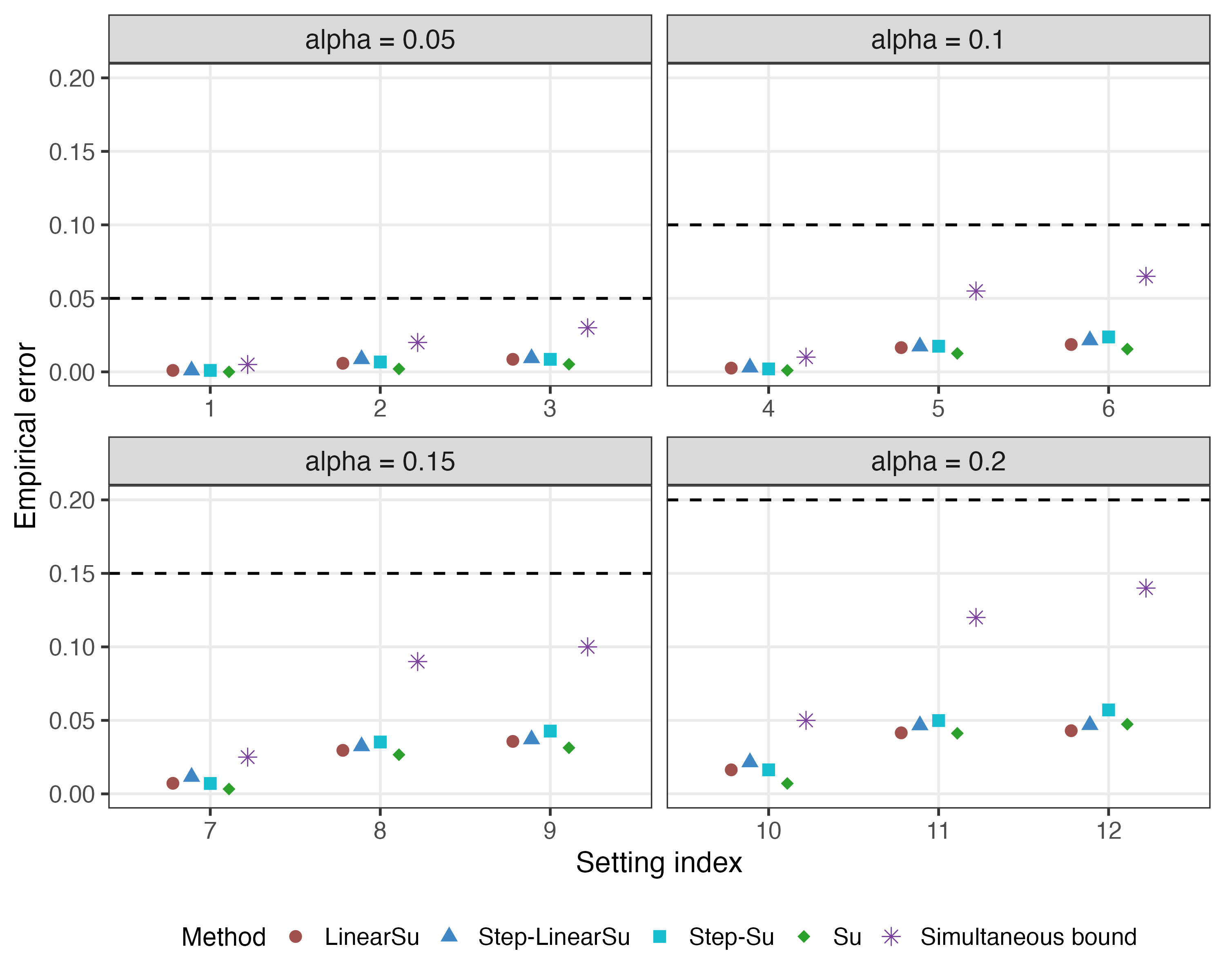}
\caption{Empirical FDR of the e-Closure-based methods with four p-to-e calibrators, along with simultaneous bound violation probabilities, across $12$ settings under varying nominal levels and intervention strengths. The settings are described in Section~\ref{sec:numericalSimu} and indexed by $1,\ldots,12$. For each nominal level $\alpha$, the three corresponding indices (in increasing order) represent intervention strengths $1$, $1.5$, and $2$, respectively.}
    \label{app:fig:empiricalErrors}
\end{figure}

The empirical FDR of the four e-Closure-based methods and the simultaneous bound violation probabilities across the $12$ simulation settings in Section~\ref{sec:numericalSimu} are shown in Figure~\ref{app:fig:empiricalErrors}. All error rates are under control.
In particular, the simultaneous bound violation probability is one minus the probability that the simultaneous guarantee holds, that is, $1 - P( |R \cap S^*| \geq t_{\alpha}(R) \text{ for any } R \subseteq [m])$
(see \eqref{simultaneous-TD-ICP}). Hence, a violation probability below the nominal level $\alpha$ implies that the simultaneous guarantee holds.

%%%%%%%%%%%%%%%%%%%%%%%%%%%%%%%%%%%%%%%%%%%%
\subsection{More simulation results on simultaneous bounds} \label{appendix:simuSimul}

\begin{figure}[htbp]
\centering
\begin{subfigure}[t]{0.7\textwidth}
    \centering
    \includegraphics[width=\textwidth]{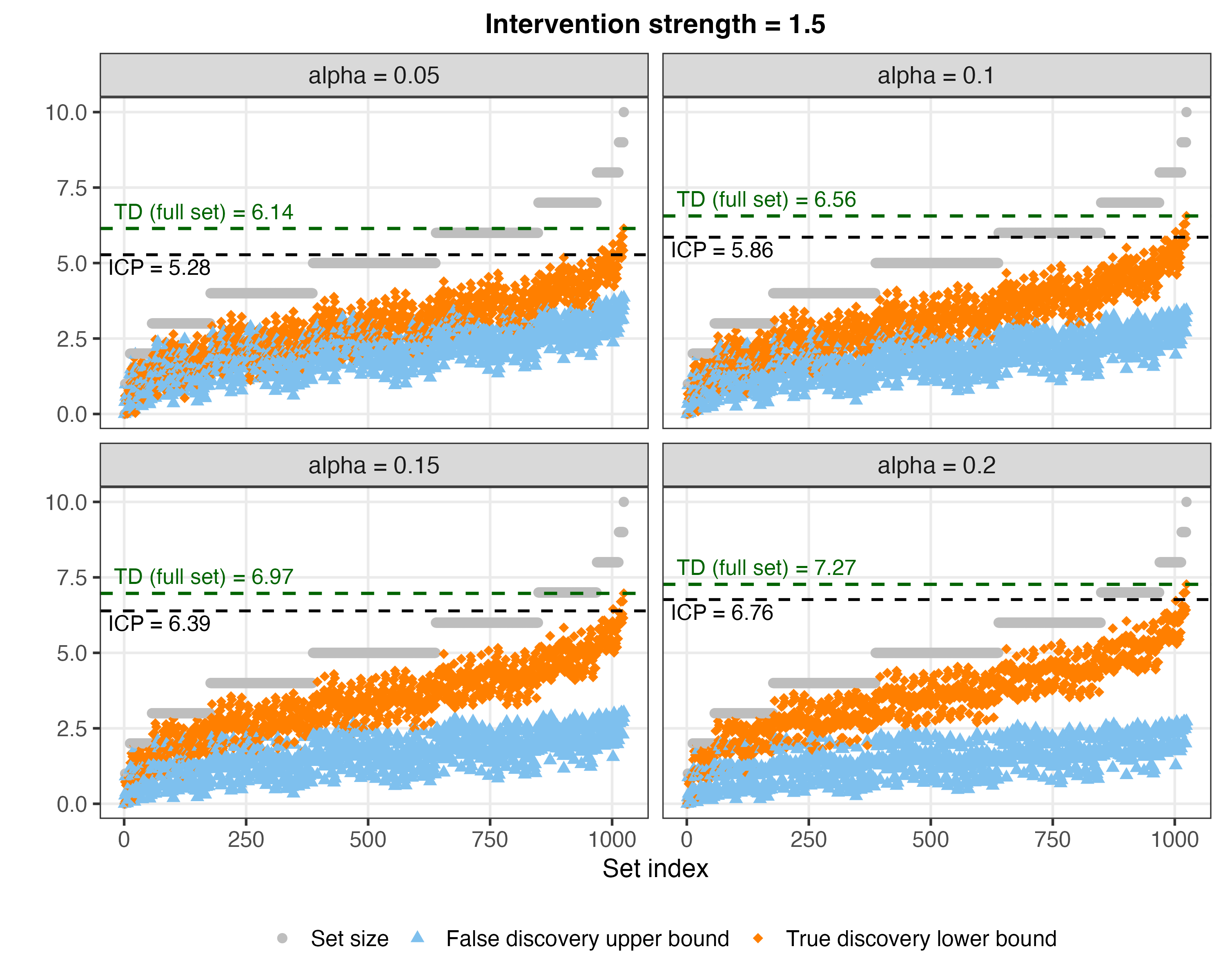}
\end{subfigure}
\hfill
\begin{subfigure}[t]{0.7\textwidth}
    \centering
    \includegraphics[width=\textwidth]{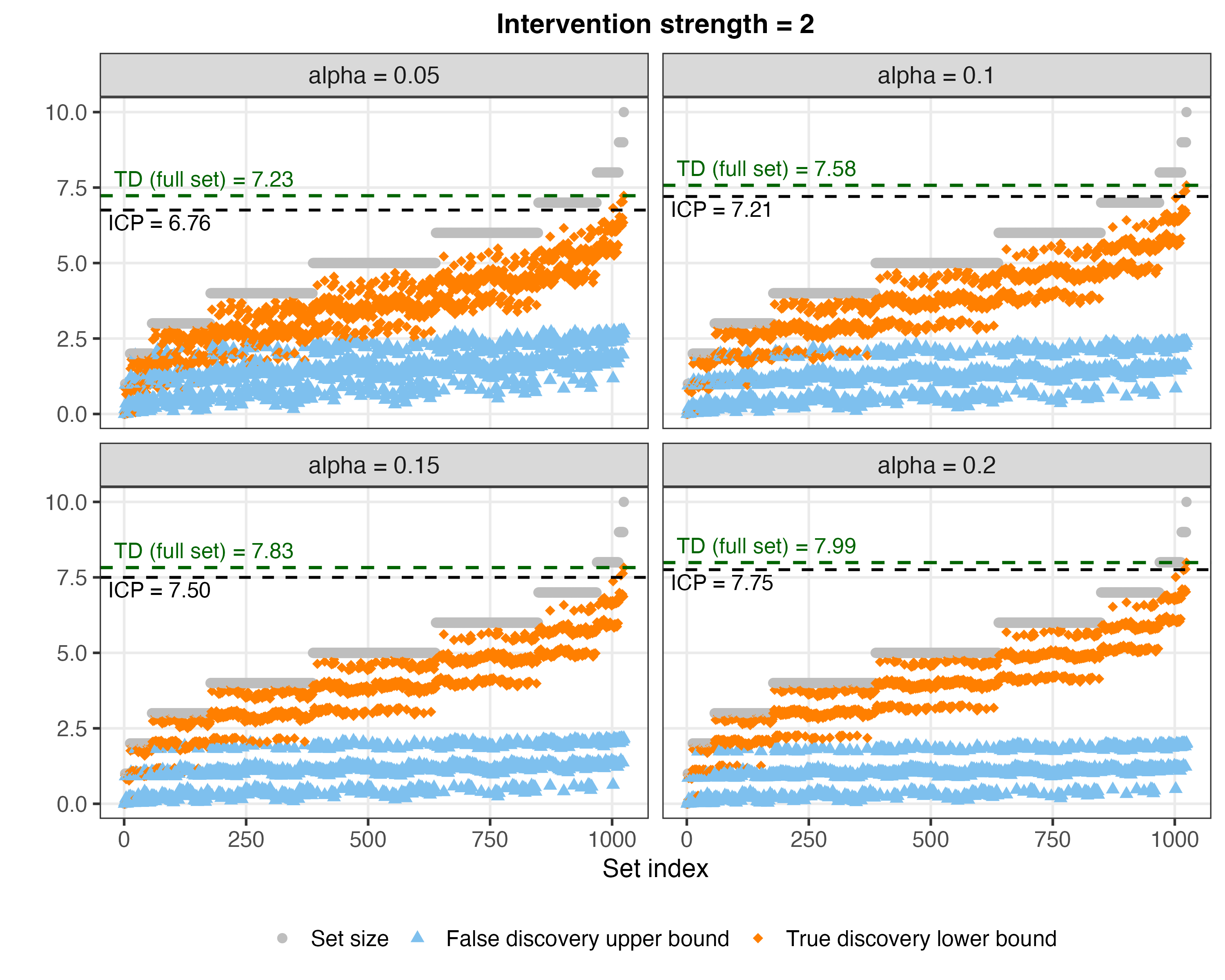}
\end{subfigure}
\caption{Average simultaneous false discovery upper bounds and true discovery lower bounds over all $1024$ subsets, together with set sizes, under intervention strengths $1.5$ (upper four plots) and $2$ (lower four plots) across different nominal levels. The black dashed horizontal line indicates the average number of discoveries of the original ICP method, while the green dashed line indicates the average simultaneous true discovery lower bound for the full predictor set.}
\label{fig:simul-combined}
\end{figure}

The corresponding simultaneous bound results for settings with intervention strengths $1.5$ and $2$ are shown in Figure~\ref{fig:simul-combined}.

\end{document}